\documentclass[12pt]{article}
\usepackage{epsfig,amssymb}
\usepackage{latexsym}

\newcommand{\bq}{\begin{equation}}
\newcommand{\eq}{\end{equation}}
\newcommand{\bqa}{\begin{eqnarray}}
\newcommand{\eqa}{\end{eqnarray}}
\newcommand{\ben}{\begin{enumerate}}
\newcommand{\een}{\end{enumerate}}
\newcommand{\bc}{\begin{center}}
\newcommand{\ec}{\end{center}}
\newcommand{\bqb}{\begin{eqnarray*}}
\newcommand{\eqb}{\end{eqnarray*}}

\def\gsim{\gtrsim}
\def\lsim{\lesssim}

\def\pr#1#2#3{ Phys. Rev. ${\bf{#1}}$:#2 (#3)}
\def\prl#1#2#3{ Phys. Rev. Lett. ${\bf{#1}}$:#2 (#3)}

\def\prep#1#2#3{ Phys. Rep. ${\bf{#1}}$:#2 (#3)}

\def\np#1#2#3{ Nucl. Phys. ${\bf{#1}}$:#2 (#3)}

\def\epj#1#2#3{ Eur. Phys. J. ${\bf{#1}}$:#2 (#3)}
\def\ijmp#1#2#3{ Int. J. Mod. Phys. ${\bf{#1}}$:#2 (#3)}

\def\arnps#1#2#3{Ann. Rev. Nucl. Part. Sci. ${\bf{#1}}$:#2 (#3) }

\def\ie{{\it i.e.}}
\def\eg{{\it e.g. }}

\global\nulldelimiterspace = 0pt

\def\sw{s_W}
\def\cw{c_W}
\def\swd{s^2_W}

\def\mwd{m_W^2}
\def\mw{m_W}
\def\mz{m_Z}
\def\mzd{m_Z^2}

\begin{document}
\pagenumbering{arabic}
\thispagestyle{empty}
\def\thefootnote{\fnsymbol{footnote}}
\setcounter{footnote}{1}

\begin{flushright}
March 6, 2008\\
PTA/07-40\\
arXiv:0709.1789 [hep-ph]\\
corrected version\\

 \end{flushright}
\vspace{2cm}
\begin{center}
{\Large\bf Remarkable virtual SUSY effects
in $W^{\pm}$ production\\
\vspace{0.2cm}
 at high energy hadron colliders}
 \vspace{1.5cm}  \\
{\large G.J. Gounaris$^a$, J. Layssac$^b$,
and F.M. Renard$^b$}\\
\vspace{0.2cm}
$^a$Department of Theoretical Physics, Aristotle
University of Thessaloniki,\\
Gr-54124, Thessaloniki, Greece.\\
\vspace{0.2cm}
$^b$Laboratoire de Physique Th\'{e}orique et Astroparticules,
UMR 5207\\
Universit\'{e} Montpellier II,
 F-34095 Montpellier Cedex 5.\\
\end{center}

\vspace*{1.cm}
\begin{center}
{\bf Abstract}
\end{center}

We present a complete  1-loop study of the  electroweak
corrections to the process $ug\to dW^+$ in MSSM and SM.
 The occurrence of
a number of remarkable properties
in the behavior of the
helicity amplitudes at high energies is stressed,
and the crucial role of the virtual SUSY contributions in establishing
them, is emphasized.
The approach to asymptopia of these amplitudes is discussed,
comparing the effects of the logarithmic and  constant contributions
to the mass suppressed ones, which are relevant  at lower energies.
Applying  crossing to $ug\to d W^+$, we  obtain  all subprocesses  needed
for the 1-loop electroweak corrections
to  $W^\pm$-production at LHC.
The SUSY model dependence of such a production is then studied,
and illustrations are given
for the transverse $W^{\pm}$ momentum distribution,
as well as the angular
distribution in the subprocess center of mass.

\vspace{0.5cm}

\def\thefootnote{\arabic{footnote}}
\setcounter{footnote}{0}
\clearpage

\section{Introduction}

The general properties  of the  virtual
supersymmetric (SUSY) electroweak corrections to the  amplitude
of any process at high energies have  already been
identified  in the literature \cite{MSSMrules}.
In particular, precise rules for all
logarithmic contributions have been established, completing
 those applying to  the Standard Model (SM)
case \cite{SMrules}. These rules provide simple and clear asymptotic tests of the
SUSY gauge and Yukawa couplings, and several applications have
been given for $e^-e^+$ and hadron colliders. \\

Moreover it has been shown in \cite{heli}, that
for any gauge supersymmetric theory, the helicity
amplitudes $F_{\lambda_a \lambda_b \lambda_c \lambda_d} $
 for any two-body processes
\bq
a_{\lambda_a}+b_{\lambda_b} \to c_{\lambda_c}+d_{\lambda_d } ~~,
\label{gen-proc}
\eq
  at fixed  angles and very high energies,
  must satisfy conservation of total helicity.
Here   $(a,b,c,d)$ denote  fermions, gauge bosons
or scalar particles, and $(\lambda_a, \lambda_b, ...)$  describe their
 helicities.
This means that at energies much higher
than all masses in the theory,
only the helicity amplitudes  obeying
\bq
\lambda_a+\lambda_b= \lambda_c + \lambda_d  ~, \label{HC-rule}
\eq
may acquire non-vanishing values.
The validity of (\ref{HC-rule}),  to all orders in any
softly broken  supersymmetric extension of SM, like \eg
MSSM,  is referred to as the  Helicity Conservation (HC) rule,
 and its general proof  has been presented in \cite{heli}.

In the non supersymmetric  standard model (SM),
HC is also  approximately correct, to 1-loop leading logarithmic
accuracy.  In such a  case, where all particles in (\ref{gen-proc})
are assumed to be  ordinary  SM ones,
the asymptotically dominant amplitudes
 should  also obey (\ref{HC-rule}); but the subdominant ones,
 which violate (\ref{HC-rule}), should asymptotically
 tend to (possibly non-vanishing) constants.

As emphasized  in \cite{heli}, the validity
of the HC  rule is particularly
tricky when  some of the participating particles
are gauge bosons; because then  large cancellations
among the various diagrams are needed for establishing it.
Moreover, the general  proof  is based on neglecting all masses
  and  the electroweak breaking scale, at asymptotic energies \cite{heli}.
  It is therefore interesting to check the  HC validity at specific complete
  1-loop calculations, in order   to be sure that no asymptotically
  non-vanishing terms, involving \eg ratio of masses, violate it.\\

Such an example is given by
the process  $ug\to d W^+$  considered here.
When neglecting the light quark masses for this process,
$u$ and $d$ quarks  always
carry negative helicities, so that  the helicity conservation  property
(\ref{HC-rule}), effectively  refers only
to the helicities of the incoming $g$ and outgoing
$W$. Thus for $ug\to d W^+$,  the  asymptotically dominant amplitudes
 determined by HC  actually are
\bq
F_{\lambda_u \lambda_g \lambda_d \lambda_W}= F_{----}~, ~  F_{-+-+}~,
\label{GBHC-amp}
\eq
which we call   gauge boson helicity conserving  (GBHC)
amplitudes. \\

The first purpose of the present paper is to explore how such
high energy and fixed angle properties  for  the helicity amplitudes
are  generated in an exact
one-loop electroweak computation, in either SM or MSSM;
and how these   asymptotic
features are corrected at lower energies
by sub-leading  contributions.

Having achieved this, the second purpose is to look at the electroweak
corrections to $W^\pm$+ jet production at the large high energy
hadron collider LHC.
Provided infrared effects are appropriately factored
out\footnote{We return to this point below.},
the relevant subprocesses  are
$qg\to q'W$, $\bar q g \to\bar q' W$ and $q\bar q'\to Wg$
 \cite{LHC}.

For such processes, QCD corrections have been carefully considered since
a long time in \cite{QCDW}, and  more recently for  the case
of the large transverse momentum  distribution  \cite{Kidonakis}.
Electroweak corrections to  the $W^\pm$ transverse momentum
$p_T$ distribution  in  SM, have been recently discussed
by K\"uhn et al. \cite{kuhnW} and by   Hollik at al \cite{HKK},
where  infrared corrections have also been included, which
necessitates considering  the  direct  photon emission,
in addition to $W$+jet production.

At high $p_T$ at LHC,  the
SM electroweak corrections turn out to be large,
due to the occurrence of single and quadratic logarithmic
effects, as expected from the aforementioned
asymptotic rules \cite{SMrules}.
In the studies of \cite{kuhnW, HKK} though, no attention had been paid to the
 behavior of the specific helicity amplitudes
and  the  SUSY contribution to them.

Consequently, as already mentioned,
these are the aspects, on which we concentrate
in the present paper.
In more detail, we study how  the complete
one loop results  for the various  $ug \to dW^+$ helicity amplitudes
match  at high energy with the asymptotic rules established in
\cite{MSSMrules, heli}, thus  assessing     the
importance of the subleading terms at LHC  energies.

The outcome is  that  supersymmetry indeed  plays
a crucial   role in establishing the gauge boson helicity conservation.
Particularly for the gauge boson helicity violating (GBHV) amplitudes
\bq
F_{---+}~,~ F_{---0}~, ~ F_{-+--}~, ~ F_{-+-0} ~, \label{GBHV-amp}
\eq
which violate (\ref{HC-rule}), it is striking  to see
how the cancellation between the standard and supersymmetric loop contributions
is realized,  enforcing the vanishing the GBHV
amplitudes at  high energies. In other words,
in a high energy  expansion of these amplitudes in MSSM,
not only the logarithmic terms cancel out, but also the tiny
 "constant"   contributions.\\

We add here that the  $ug\to d W^+$   processes
has been chosen because of its theoretical  simplicity; not necessarily
because of its   best observability, or of its  largest
SUSY effects. It  only constitutes a simple
toy for studying the supersymmetric effects on
the helicity amplitudes.
The properties we find  should be instructive
and indicative of those expected  for other types of processes
accessible for hadron, lepton or photon colliders.
 We hope  to undertake such studies in the future.\\

The contents of the next Sections is the following.
In Sect. 2 we consider the basic $ug\to dW^+$ process,
defining  the kinematics and  helicity
amplitudes and  classifying the various one loop diagrams.
In Sect. 3 we present  the detail behavior of the
various  GBHC and GBHV amplitudes at one loop
in  SM and  MSSM. The importance
of the various high energy components (leading logs, constant terms,
mass-suppressed  terms) and the role of SUSY, are discussed
by considering several benchmark models of the constrained  MSSM type.

Using then $ug\to dW^+$  and the processes  related to it
by crossing, as well as the appropriate parton distribution functions  (PDF)
\cite{Durham},  we  present in Sect.4
the  transverse momentum distribution  for $W^{\pm}$
production in association with a jet  at  LHC.
In addition to it,  the $W^\pm$ angular distribution
in the subprocess center of mass is also shown.
The  observability
of these supersymmetry properties is  briefly discussed.
Finally, in Sect.5 we give  our conclusions and suggest some
further applications.\\

\section{ One loop electroweak amplitudes for $u g\to d W^+$.}

The momenta and helicities in this process are defined by
\bq
u(p_u, \lambda_u)+g(p_g, \lambda_g) \to d(p_d, \lambda_d)+ W^+(p_W, \lambda_W) ~~,
\label{ugdW-proc}
\eq
and the corresponding helicity amplitudes are denoted as
$F_{\lambda_u \lambda_g \lambda_d\lambda_W}$.
Neglecting  the $(u,d)$-quark masses  and remembering that
the $W$-quark coupling is purely left-handed implying
  $\lambda_u=\lambda_d=-1/2$,   while the gluon and $W$
  helicities can be  $(\lambda_g=\pm 1),~ (\lambda_W=\pm 1,0)$,
we end up with only 6 possibly non-vanishing helicity
amplitudes, which are\footnote{The sign of the amplitudes $F$, relative to
the $S$-matrix, is defined through $S=iF$. The  sign of the gauge
couplings are fixed by  writing
the covariant derivative acting on the left-quarks as
\[
D_\mu = \partial_\mu -i g_s \frac{\lambda^a}{2} G_\mu^a
+ i g \frac{\vec \tau}{2} \cdot \vec W_\mu +i g' Y B_\mu
~~,\]
where $G^a_\mu$ is the gluon field. Note that the convention for $g_s$ is opposite
to that for $g$ and $g'$.}
\bq
F_{----}~, ~  F_{-+-+} ~,~ F_{---+}~,~ F_{---0}~,~ F_{-+--}~,~ F_{-+-0} ~~.
\label{heli-amp}
\eq
 As it has already been mentioned immediately after (\ref{GBHC-amp}),
 the first two of these amplitudes satisfy the HC rule (\ref{HC-rule})
and are called GBHC.
The remaining amplitudes, which    have already appeared
in  (\ref{GBHV-amp}), are  called GBHV. Since they
violate HC, they must  vanish asymptotically, and   as will see below,
they are usually very small, also for  LHC energies.
It is convenient for the discussion below to separate them in two pairs:
namely $(F_{---+},F_{---0})$ referred to as GBHV1 amplitudes,
and $(F_{-+--}~,~ F_{-+-0})$ referred to as GBHV2.

\begin{figure}[p]
\vspace*{-1cm}
\[
\epsfig{file=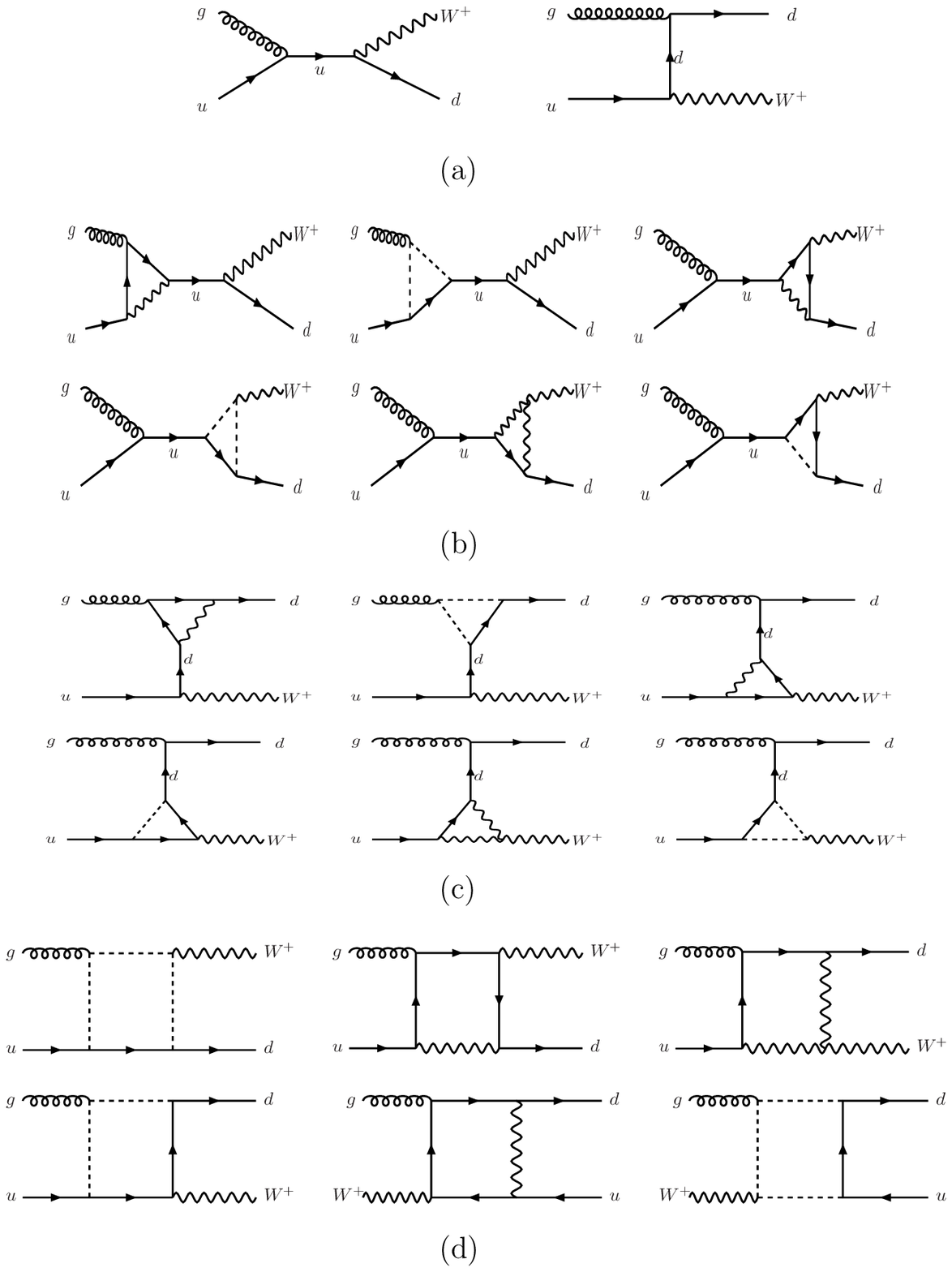,width=14cm, height=18.cm}
\]
\caption[1]{Independent Diagrams used for calculating  the $ug\to dW$
helicity amplitudes. They consist of  tree diagrams (a),
 s-channel 1-loop triangles (b),
  u-channel 1-loop triangles (c),  and  boxes (d). Full, broken and
  wavy  lines describe respectively fermionic,   scalar and gauge particles. }
\label{Diagrams}
\end{figure}

Defining  the kinematical variables
\bqa
&& s=(p_g+p_u)^2=(p_W+p_d)^2~~,~~ \beta'=1-{m^2_W\over s} ~, \nonumber \\
&& u=(p_d-p_g)^2=(p_u-p_W)^2=- {s\beta'\over2}(1+\cos\theta) ~, \nonumber \\
&& t=(p_g-p_W)^2=(p_u-p_d)^2=- {s\beta'\over2}(1-\cos\theta)~~,    \label{kin-var}
\eqa
we first turn to  the contribution of
the Born diagrams in Fig.\ref{Diagrams}a,  containing u and d quark exchanges,
in the s- and u-channel respectively. These  affect  the GBHC amplitudes
$F_{----}$, $F_{-+-+}$, and the GBHV2 ones $F_{-+--}$,  $F_{-+-0}$.
For  transverse ($\lambda_W=\pm 1$)
and longitudinal ($\lambda_W=0$)  $W^+$, these amplitudes
 are given respectively by
\bqa
F^{\rm Born}_{-,\lambda_g,-,\lambda_W}&=&\left ( \frac{ \lambda^a}{2}\right )
{eg_s\sqrt{\beta'}\over2\sqrt{2}s_W}
\cos{\theta\over2} \Bigg \{\Big [(1-\lambda_g)(1-\lambda_W)\Big ]\nonumber\\
&&+{1\over\beta'}
\Big [(1+\lambda_g)(1+\lambda_W)
+{t\over u}\Big (1-\lambda_g(1-{2m^2_W\over s})\Big )(1-\lambda_W)\Big ] \Bigg \}
 ~, \label{F-BornT} \\
F^{\rm Born}_{-,\lambda_g,-,0} &= & \left ( \frac{\lambda^a}{2}\right )
{eg_s  \over 2 \mw s_W} ~ \sin{\theta\over2}
\Bigg \{ -\sqrt{s \beta'} (1-\lambda_g)+\frac{\mw^2 (3\lambda_g+1)+s(1-\lambda_g)}
{\sqrt{s\beta'}} \Bigg \} ~, \nonumber \\
&= & \left ( \frac{\lambda^a}{2}\right )
\frac{e g_s }{\sw}\frac{\mw }{ \sqrt{s\beta'}}(1+\lambda_g) \sin{\theta\over2}~.
 \label{F-Born0}
\eqa
In (\ref{F-BornT}, \ref{F-Born0}) the factor $\lambda^a/2$ describes
the color matrices acting between the initial $u$ and final $d$ quark,
while  the first and second terms
within the curly brackets come respectively from the   s- and u-channel
diagrams in Fig.\ref{Diagrams}a.

We also note that since the amplitude $F_{-,\lambda_g,-,0}$,
given in  (\ref{F-Born0}),
 can never satisfy the HC rule (\ref{HC-rule}), it has to vanish asymptotically,
 to any order in perturbation theory. The last expression in (\ref{F-Born0}),
 is simply a tree order realization of this. \\

At the 1-loop level, the  amplitudes
in (\ref{F-BornT}, \ref{F-Born0})
receive also contributions from   the  counter terms induced by
the renormalization of the  external particle fields and coupling constants,
and determined
by various gauge, $u$- and $d$-quark self energy diagrams.
As input parameters in our renormalization scheme, we use the $W$ and $Z$ masses,
through which the cosine of the Weinberg angle is also fixed; while the fine
structure constant  $\alpha$ is defined through the
Thompson limit  \cite{Hollikscheme}.

In SM, the aforementioned    $u$- and $d$-quark self-energies
 are induced by the quark-gauge boson bubbles, while the Higgs
and Goldstone bosons effects  are negligible. The SM
gauge  self-energies come  from gauge, Higgs and fermion loops.

Correspondingly, the main SUSY contribution to the quark self energies,
consists of  the   squark-gaugino bubbles,
while  the additional SUSY Higgs bosons effects are again negligible.
For the gauge self energies though, the SUSY contribution arises from
gaugino,  higgsino and sfermion loops,
as well as the effects  related to the two-doublet
Higgs fields.

Including these counter term (c.t.) contributions  to the above Born amplitudes,
modifies them as
\bqa
F^{\rm Born+c.t.}_{-,\lambda_g,-,\lambda_W}&=&\left ( \frac{ \lambda^a}{2}\right )
{eg_s\sqrt{\beta'}\over2\sqrt{2}s_W}
\cos{\theta\over2} \Bigg \{\Big [ (1+\delta_s) (1-\lambda_g)(1-\lambda_W)\Big ]\nonumber\\
&&+{(1+\delta_u) \over\beta'}
\Big [(1+\lambda_g)(1+\lambda_W)
+{t\over u}\Big (1-\lambda_g(1-{2m^2_W\over s})\Big )(1-\lambda_W)\Big ] \Bigg \} ~,
\label{F-BornT-ct} \\
F^{\rm Born+c.t.}_{-,\lambda_g,-,0} &= & \left ( \frac{\lambda^a}{2}\right )
{eg_s  \over 2 \mw s_W} ~ \sin{\theta\over2}
\Bigg \{ -\sqrt{s\beta'} (1-\lambda_g)(1+\delta_s)
\nonumber \\
&+ & \frac{\mw^2 (3\lambda_g+1)+s(1-\lambda_g)}
{\sqrt{s\beta'}}(1+\delta_u) \Bigg \}, \nonumber \\
&= & \left ( \frac{\lambda^a}{2}\right )
{eg_s \mw \over  s_W \sqrt{s\beta'}} ~ \sin{\theta\over2}
\Bigg \{ (1+\lambda_g)(1+\bar\delta) \nonumber \\
  & - & \frac{\Sigma^d_L(u)(1+3\lambda_g)}{2}-\frac{\Sigma^u_L(s)(1-\lambda_g)}{2}
   +  \frac{s(1-\lambda_g)[ \Sigma^u_L(s)-\Sigma^d_L(u)]}{2\mw^2} \Bigg \} ~,
   \label{F-Born0-ct}
\eqa
where
\bqa
&& \delta_s =  \bar \delta - \Sigma^u_L(s) ~~, ~~
\delta_u =  \bar \delta - \Sigma^d_L(u) ~~, \label{u-d-self} \\
&& \bar \delta  =\delta Z_1^W-\delta Z_W+{1\over2}\delta\Psi_W
+{1\over2}\delta Z^d_L +{1\over2}\delta Z^u_L  ~. \label{ct-terms}
\eqa
In   (\ref{ct-terms}), the first three terms in the r.h.s.
come from the renormalization of the  gauge couplings, masses
and wave functions,  through
\bqa
&& \delta Z_W = - \Sigma^{T'}_{\gamma\gamma}(0)
+2{c_W\over s_W m^2_Z}\Sigma^T_{\gamma Z}(0)
+{c^2_W\over s^2_W}\left [{\delta m^2_Z\over m^2_Z} -
{\delta m^2_W\over m^2_W}\right ]~, \nonumber \\
&& \delta m^2_W=Re\Sigma^T_{WW}(m^2_W)~~,~~\delta m^2_Z=Re\Sigma^T_{ZZ}(m^2_Z)~,
\nonumber \\
&& \delta Z_1^W-\delta Z_W =  {\Sigma^T_{\gamma Z}(0)\over s_Wc_W M^2_Z}~,
~ \Sigma^T_{\gamma Z}(0) = -\frac{\alpha}{2\pi}\frac{\mwd}{\sw\cw}
\left[ \Delta -\ln\frac{\mwd}{\mu^2} \right ]~, \label{delta-W} \\
&&
\delta\Psi_W=-Re\hat\Sigma^{T'}_{WW}(m^2_W)=
-\{Re\Sigma^{T'}_{WW}(m^2_W) + \delta Z_W \}~~, \label{delta-psiW}
\eqa
where $\Delta$ describes the usual ultraviolet contribution, in  dimensional
regularization. The needed gauge self energies in SM and SUSY may be found \eg
in the  appendices of \cite{eeV1V2-long},
expressed in terms of Passarino-Veltman (PV)
functions \cite{Veltman}.

Since, as discussed below, the infrared
divergencies are always regularized by  a non-vanishing
"photon mass" $m_\gamma$,  this "photon mass"  must be inserted into
the various $B_j$  functions  taken from   \cite{eeV1V2-long}.
In addition to this,
the quantity $\frac{\alpha}{2\pi}m_\gamma^2\Delta $
must be added to the r.h.s. of
the expression (C.18) of \cite{eeV1V2-long}.

We also note that  the last two  terms in the r.h.s of
(\ref{ct-terms}), as well as  (\ref{u-d-self}),
come from the external quark wave functions
and the  self energies of  the intermediate quarks in Fig.\ref{Diagrams}a.
These can also be obtained  from the   appendices of \cite{eeV1V2-long}.

Finally, the validity of the HC rule for the amplitude
(\ref{F-Born0-ct}), ensures that the last term within its curly brackets,
which depends on  quark self energies,
 must be cancelled at high energies, by some triangular
and box contributions.  \\

Having finished with the c.t. contributions to $ug\to d W^+$,
  we now turn to the
triangular and box contributions from Fig.\ref{Diagrams}.

The topologies of the triangular graphs consist of the left and
right s-channel  triangles appearing in Fig.\ref{Diagrams}b, and
 the up  and  down u-channel triangles shown Fig.\ref{Diagrams}c.
 The full, broken and wavy  lines in this figure, describe respectively the various
 fermionic, scalar and gauge particles in SM or MSSM.
We have checked explicitly that the ultraviolet divergences contained in these
graphs, cancel exactly those induced by the counter
terms in (\ref{F-BornT-ct}, \ref{F-Born0-ct}).

The    boxes  for  $ug\to dW$ are indicated
in Fig.\ref{Diagrams}d. The first two boxes are   direct boxes,
the next two boxes   are the crossed ones, and the final two
are the twisted boxes. All possible gauge, fermion and
scalar exchanges should be  taken into account, in both SM and MSSM.

Using  (\ref{F-BornT-ct}, \ref{F-Born0-ct}) and the triangular
and box graphs mentioned above, we express   the
 complete 1-loop electroweak amplitudes for $ug\to d W^+$
in the form
\bq
F^{\rm 1-loop} = \left ( \frac{ \lambda^a}{2}\right )
\sum_{i=1}^{10} N_i(s,t,u)~ \bar u_d K_iP_L  u_u ~, \label{F-complete}
\eq
where the factor $\lambda^a/2$ describe, as before,
the color matrix elements  between the initial $u$ and the final
$d$ quark. In  (\ref{F-complete}),
  $K_i$ is a set of 10 invariant forms constructed by Dirac
matrices, gluon and $W$ polarization vectors and external
momenta, acting between the $u$ and $d$ quark Dirac wave functions.
The helicity amplitudes are  computed from them,   using
appropriate Dirac wave functions. Finally,
$ N_i(s,t,u)$ are the corresponding  scalar quantities calculated from
the various diagrams in terms of  PV functions, and depending
on $(s,t,u)$ and the couplings and masses. \\

We have already mentioned the  ($\alpha, m_Z, m_W$)-parameters,
used as input in our scheme. There is also in the
$W$ counter term a slight dependence in the Higgs mass.
In addition to them, the SUSY diagrams
involve contributions from the chargino, neutralino and the squark masses,
 as well as their mixing.
 The SUSY effect is illustrated in the next Section by considering
three particular constrained MSSM benchmarks presented in Table 1.
\begin{table}[h]
\begin{center}
{ Table 1: Input  parameters at the grand scale,
for three  constrained MSSM benchmark models. We always have  $\mu>0$.
All dimensional parameters are in GeV. }\\
  \vspace*{0.3cm}
\begin{small}
\begin{tabular}{||c|c|c|c||}
\hline \hline
  &  BBSSW & $SPS1a'$ & light SUSY      \\ \hline
 $m_{1/2}$ &900  &   250 & 50   \\
 $m_0$ & 4716  & 70 & 60   \\
 $A_0$ &0 & -300  & 0   \\
$\tan\beta$ & 30  & 10  & 10  \\
  \hline \hline
$M_{SUSY}$ & 700  & 350  & 40  \\
  \hline \hline
\end{tabular}
 \end{small}
\end{center}
\end{table}

The first of these benchmarks
is a "heavy scale" model called here BBSSW,
which has been suggested by  \cite{Baer} under  the name FP9.
It is  a focus point scenario, analogous to mSP1 of \cite{Nath},
and consistent with all present experimental information\footnote{As is well known,
the consistency of a constrained focus point MSSM model depends
sensitively on  the top mass.
In the present model $m_t=175$GeV has been used in \cite{Baer}.
The results of the present paper though,
are not sensitive to the top mass.}.
The $M_{SUSY}$ parameter in Table 1, is discussed below.

The second
is the "medium scale" model SPS1a',  advocated in \cite{SPA}. It is
very close to the mSP7 model of \cite{Nath},
and it is also contained  in  \cite{OBuchmueller}. It
is consistent with all present knowledge.

Finally, the  "light scale" model appearing
in the last column of Table 1, is
already experimentally excluded. But it is nevertheless
useful for the present discussion, since it  gives
a picture of the two-body amplitudes at energies much larger
than all SUSY masses. This is particularly useful for showing how the
various GBHC and the GBHV amplitudes reach their asymptotic limits,
as the SUSY masses become much smaller than the available energy. \\

As already mentioned, to avoid the infrared
divergences we impose $m_{\gamma}=\mz$.
A similar choice has also been  made in \cite{MSSMrules},
when considering the properties of the Sudakov logs.
As pointed out by   Melles,  this has the advantage of treating
the $\gamma$, $Z$ and $W^{\pm}$ contributions on  the same footing,
for  $\sqrt{s}\gg \mz$; thus preserving
the $SU(2)\otimes U(1)$ gauge symmetry \cite{mgamma}.

In this  scheme, it is perfectly consistent
to  restrict to  the $W$+jet production at LHC,
without including the  direct hard photon emission
\cite{kuhnW, HKK}. We explicitly assume here, that this
is experimentally possible\footnote{A similar spirit is
followed in \cite{Kidonakis}.}.
On the other hand, the direct photoproduction of soft photons
  need not be included in our scheme,
since it is part of the complementary pure QED contribution
defined as the infrared finite quantity formed
by  combining    the real photon emission
with $m_{\gamma}=\lambda$, using the   appropriate
experimental cuts  \cite{kuhnW, HKK}, and further adding
the difference between
the  virtual photon exchanges for $m_{\gamma}=\lambda$
and $m_{\gamma}=\mz$. \\

Using this, we study  the properties
of the electroweak corrections in SM and  SUSY.
The  genuine  SM corrections  arise from the  quark,
 $Z$, $W^{\pm}$ and $\gamma$ exchanges, with $m_{\gamma}=\mz$;
while   by SUSY corrections are induced by the additional contributions involving
sfermion, chargino or   neutralino exchanges.
The sum of these SM and SUSY  contributions,  constitute the complete
"MSSM electroweak corrections". \\

A FORTRAN code in which the six helicity amplitudes are
computed using as input the needed MSSM parameters
is available at the
site\footnote{A factor $\lambda^a/2$ has been removed from the
amplitudes given in the code.
All other conventions are as in this paper.}   \cite{code}.

\subsection {High energy behavior of the $ug\to d W^+$ amplitudes}
For $s,|t|, |u| \gg \mw^2$,  the dominant  amplitudes are of course
the GBHC ones, $F_{----}$ and $F_{-+-+}$ \cite{heli}. At the Born-approximation
these tend to the limits
\bqa
F^{\rm Born}_{----} & \to &
{eg_s\over\sqrt{2}s_W} \left ({\lambda^a\over2} \right )
{2\over \cos{\theta\over2}} ~~, \nonumber \\
F^{\rm Born}_{-+-+} & \to  & {eg_s\over\sqrt{2}\sw} \left ({\lambda^a\over2}\right )
{2 \cos{\theta\over2}} ~~, \label{F-Born-HC-as}
\eqa
while the GBHV2 ones are vanishing as
\bqa
F^{\rm Born}_{-+--} & \to & {\sqrt{2} eg_s\over \sw}
\left ({\lambda^a\over2}\right )
\left ({t \over u} \right ) \frac{\mw^2}{s}  \cos{\theta\over2}
\simeq 0~~, \nonumber  \\
F^{\rm Born}_{-+-0} & \to & {2 eg_s\over \sw}
\left ({\lambda^a\over2}\right )  \frac{\mw}{\sqrt{s}} \sin{\theta\over2}
\simeq 0~~,  \label{F-Born-HV2-as}
\eqa
and the GBHV1 amplitudes  satisfy
\bq
F^{\rm Born}_{---+}=F^{\rm Born}_{---0}=0 ~~.  \label{F-Born-HV1-as}
\eq
Here, (\ref{F-Born-HC-as}, \ref{F-Born-HV2-as})  come from taking
the high energy limit in (\ref{F-BornT}, \ref{F-Born0}),
while (\ref{F-Born-HV1-as}) is a
consequence of neglecting the  $u$ and $d$ quark masses.
As it should be,
 (\ref{F-Born-HC-as}, \ref{F-Born-HV2-as}, \ref{F-Born-HV1-as} )
respect the HC rule at asymptotic energies \cite{heli}.

At one loop, for $(s,|t|,|u|)$ much larger than all masses exchanged
in the diagrams, the real parts of the GBHC amplitudes, including
the  leading logarithmic  corrections,  are given by
\bqa
 ReF_{----} &\simeq &  {eg_s\over\sqrt{2}s_W}\left ({\lambda^a\over2}\right )
{2\over\cos{\theta\over2}}
\Bigg \{1+{\alpha\over4\pi}{(1+26c^2_W)\over36s^2_Wc^2_W}
\Big [ 3\ln{s \over m^2_Z}-\eta \ln{s\over M^2_{SUSY}}
-\ln^2{s \over m^2_Z} \Big ]
\nonumber \\
& -& {\alpha\over4\pi s^2_W} \ln^2 {s \over m^2_W}
- {\alpha\over4\pi}\Bigg [ {(1-10c^2_W)\over36s^2_Wc^2_W}
\Big (\ln^2{-t\over m^2_Z}-\ln^2{s \over m^2_Z} \Big ) \nonumber\\
& +& {1\over2s^2_W}\Big ( \ln^2{-u\over m^2_Z}+\ln^2{-u\over m^2_W}
-\ln^2{s \over m^2_Z}-\ln^2{s \over m^2_W} \Big ) \Bigg ]
\nonumber \\
&+ &\frac{\alpha}{4\pi}  [C^{SM}_{----}+
\eta C^{SUSY}_{----}  ] \Bigg \}, \label{ReFmmmm-LL}\\
ReF_{-+-+} &\simeq & {eg_s\over\sqrt{2}s_W}\left ({\lambda^a\over2}\right )
2\cos{\theta\over2} \Bigg \{1+{\alpha\over4\pi}{(1+26c^2_W)\over36s^2_Wc^2_W}
\Big [ 3\ln{s\over m^2_Z}-\eta \ln{s\over M^2_{SUSY}}
-\ln^2{s \over m^2_Z} \Big ]
\nonumber \\
&  - & {\alpha\over4\pi s^2_W}\ln^2{-s-i\epsilon \over m^2_W}
- {\alpha\over4\pi}\Bigg [{(1-10c^2_W) \over36s^2_Wc^2_W}
\Big ( \ln^2{-t\over m^2_Z}-\ln^2{s \over m^2_Z}\Big ) \nonumber\\
& +& {1\over2s^2_W}\Big ( \ln^2{-u\over m^2_Z}+\ln^2{-u\over m^2_W}
-\ln^2{s \over m^2_Z}-\ln^2{s \over m^2_W} \Big )\Bigg]
\nonumber \\
&+ &\frac{\alpha}{4\pi}  [C^{SM}_{-+-+}+\eta C^{SUSY}_{-+-+} ]\Bigg \},
\label{ReFmpmp-LL}
\eqa
where  $\eta=0$ in SM, and  $\eta=1$ for SUSY.
 The $M_{SUSY}$ quantity,    denoting  an   average of the gaugino
and squark masses involved in the process, is given in Table 1,
 for each benchmark model.

The last terms  within the curly brackets in
 (\ref{ReFmmmm-LL}, \ref{ReFmpmp-LL}) give
 the subleading non-logarithmic  contributions, described by
 $C^{SM}_{-\mp-\mp}$ and $C^{SUSY}_{-\mp-\mp}$, which are referred to as "constant"
 contributions. These "constants"   are  energy  independent
  quantities,  possibly depending on the model, due to their dependence
  on the ratios of internal and external masses, and also on the
  angle.  Note also that
 there is a correlation between exact values of  $M_{SUSY}$ and $C^{SUSY}$;
 a change of one may be absorbed in the other.

 In the MSSM  case, when all SM and SUSY contributions
 are taken into account, we also define
\bq
C^{MSSM}_{-\mp-\mp} =C^{SM}_{-\mp-\mp}+C^{SUSY}_{-\mp-\mp} ~~. \label{CMSSM}
\eq
\begin{table}[h]
\begin{center}
{ Table 2: Angular dependence of the $C^{SM}_{-\mp-\mp}$ and
$C^{MSSM}_{-\mp-\mp}$ parameters for the three  constrained
MSSM benchmark models used here.  }\\
  \vspace*{0.3cm}
\begin{small}
\begin{tabular}{||c|c|c|c|c||}
\hline \hline
  & \multicolumn{2}{|c|}{SM } &  \multicolumn{2}{|c|}{MSSM }   \\
    \hline
$\theta $    & $C_{----}$ & $C_{-+-+}$ &  $C_{----}$ & $C_{-+-+}$ \\ \hline
    $30^o$ & 23 & 15   & 22 & 14  \\
     $60^o$ & 23  & 19 & 25  & 21 \\
      $90^o$ &  19 & 21  & 23 & 23  \\
       $150^o$ & 16 & 42   &29 & 45  \\
   \hline \hline
\end{tabular}
 \end{small}
\end{center}
\end{table}

A judicious choice of these  "constants", has been  obtained by comparing
the dominant real parts of the  exact 1-loop prediction  for SM and
the three MSSM models of Table 1,
with those from (\ref{ReFmmmm-LL}, \ref{ReFmpmp-LL}).
This  gives the results presented in Table 2, for SM and the three
benchmark MSSM models of  Table 1.
Note that the  "constants" in Table 2  look amply plausible,  when  compared to the
asymptotic expressions of the PV functions \cite{Denner-Roth, techpaper}.
And they also seem very little depending on the MSSM model. \\

Concerning  (\ref{ReFmmmm-LL}, \ref{ReFmpmp-LL}), it is important to emphasize
that the coefficient of  $\ln |s|$, which is $3$ in SM, it  is reduced to
 $3-\eta=2$ in   MSSM \cite{MSSMrules, SMrules}.
 This is a striking difference between SM and MSSM, which does not depend
 on the specific value of any parameter of supersymmetric origin.

As it can be seen from the code released in \cite{code},
the imaginary parts of GBHC amplitudes are much smaller
than the real parts for energies below the TeV-range.
At high energy they can also be approximately described by
\bqa
ImF_{----} &\simeq &  {eg_s\over\sqrt{2}s_W}\left ({\lambda^a\over2}\right )
{2\over\cos{\theta\over2}} \left \{{\alpha\over4\swd}
 \Big [\ln\frac{s}{\mzd}+\ln\frac{s}{\mwd}    \Big ]\right \}~~, \label{ImFmmmm-LL}\\
ImF_{-+-+} &\simeq & {eg_s\over\sqrt{2}s_W}\left ({\lambda^a\over2}\right )
2\cos{\theta\over2} \left \{{\alpha\over4\swd}
 \Big [\ln\frac{s}{\mzd}+\ln\frac{s}{\mwd}    \Big ]\right \}~~,  \label{ImFmpmp-LL}
\eqa
to very good accuracy. These contributions only come from the
SM part; SUSY contributions are negligible. The numerical
agreement between the above expressions and the exact
computations means that the aforementioned "constants"
are mainly real.\\

It should be now interesting to see how and in which
amplitudes the MSSM parameters (couplings, masses and mixings)
enter progressively the game
beyond the logarithmic high energy approximation,
by contributing to the successive
subleading terms; constant and  mass suppressed terms of
successive orders... .
 This is shown in the  illustrations  of  the next Section.\\

\section{One loop SUSY effects in the  $u g\to d W^+$ process}

In this Section we explore in more detail
the specific properties of the
helicity amplitudes, from threshold to "asymptotic" energies.
Using (\ref{F-BornT}, \ref{F-Born0})
and the 1-loop results described above, we show  in the
figures below, first the features of the dominant
GBHC amplitudes $F_{----},~F_{-+-+}$; and subsequently
those of  the subdominant pairs
GBHV1 $(F_{---+},F_{---0})$,
and GBHV2 $(F_{-+-+},F_{-+-0})$.\\

We will concentrate our discussions and show illustrations
only for the real parts of the helicity amplitudes (although
the code produces both real and imaginary parts). The reason
is that imaginary parts are usually much smaller, except
for energy values close to thresholds for intermediate
processes. This is particularly true for the
for the GBHC and GBHV2 amplitudes which receive purely
real Born contributions. Only for GBHV1 amplitudes, which
receive no Born contribution, the imaginary parts can be
comparable to the real parts close to these thresholds.
But these amplitudes are very small and quickly decrease with the energy.

\subsection{Features of the GBHC amplitudes $F_{----},~F_{-+-+}$. }

We start with the dominant real parts of the helicity conserving amplitudes
GBHC calculated  in the  Born and 1-loop
approximation, in SM and the three  MSSM models of Table 1.
    Fig.\ref{HC-energy-fig}a shows the energy dependence for
   quark-gluon c.m. energies
 $\sqrt{s}\leq 0.6 $TeV, while   Fig.\ref{HC-energy-fig}b concerns the region
  $\sqrt{s}\lsim  20 $TeV. The c.m. scattering angle in the figures
 is fixed at $\theta =60^o$. The coefficient
 $\lambda^a/2$ is always  factored out.

As seen in  Fig.\ref{HC-energy-fig}a,b,
the Born amplitudes,  become constant for $\sqrt{s}\gsim 0.3$TeV.

Including the SM 1-loop corrections, a
positive effect arises below 0.4 TeV, which at higher energies
becomes negative and  increasing in magnitude,
in agreement with the log rules in (\ref{ReFmmmm-LL}, \ref{ReFmpmp-LL}).

When the SUSY corrections contained in the MSSM models of  Table 1 are included,
the amplitudes are further reduced compared to their SM values,
with the reduction  becoming stronger as we move  from BBSSW, to $SPS1a'$ and
"light SUSY". This is understandable
 on the basis of (\ref{ReFmmmm-LL}, \ref{ReFmpmp-LL}),
 since   $M_{SUSY}$ decreases in this direction.\\

In fact, (\ref{ReFmmmm-LL}, \ref{ReFmpmp-LL}) describe very accurately
the 1-loop SM and MSSM results for $\sqrt{s}\gsim 0.5$TeV,
provided we use  the "constants" given in Table 2
and the $M_{SUSY}$-values of Table 1,
for all our benchmark models. To assess the accuracy of
(\ref{ReFmmmm-LL}, \ref{ReFmpmp-LL}), we compare them to the exact
1-loop results for SM and $SPS1a'$ in
Fig.\ref{HC-asym-fig}a and \ref{HC-asym-fig}b respectively,
using   $\theta =60^o$.
A similar accuracy is also obtained for the other
two benchmark models we have considered; BBSSW and "light SUSY".

The angular distributions for the GBHC amplitudes
at $\sqrt{s}=0.5$TeV and $\sqrt{s}=4$TeV,
are shown in Figs.\ref{HC-angle-fig}a and \ref{HC-angle-fig}b respectively.
Their  shapes are almost identical  for SM and all MSSM models considered,
and  very similar to the Born ones.\\

Quite accurate are also the expressions
(\ref{ImFmmmm-LL}, \ref{ImFmpmp-LL}) for the imaginary parts
of the GBHC amplitudes, which of course are much smaller
than the real parts.\\

\subsection{Features of the GBHV1 amplitudes $F_{---+},F_{---0}$.}

The GBHV1 pair of amplitudes $F_{---+},F_{---0}$ are shown in
Fig.\ref{HV1-energy-fig}a,b,  for $\sqrt{s}\leq 0.6$TeV
and for $\sqrt{s}\lsim 20$TeV respectively,
and the same value  $\theta=60^o$.
Since there is no Born contribution to these amplitudes,
the figures show only  the 1-loop prediction for
 SM and the three benchmark MSSM models of Table 1.
The structure observed around 0.2 TeV in the light model,
is due to a SUSY threshold effect to which no attention
should be paid, as this model is already experimentally
excluded.
 Above 0.2TeV, these amplitudes are much smaller than GBHC ones;
 compare  to Fig.\ref{HC-energy-fig}a,b.

According to the HC rule \cite{heli},
 both these amplitudes should vanish at very high energies  in MSSM,
 while in SM they may tend to  non-vanishing constant values.
Such a non-vanishing limit for  $F_{---+}$ in SM, may be seen in
Fig.\ref{HV1-energy-fig}b.

For the MSSM cases though, $F_{---+}$ tends to vanish at high energies,
with these energies strongly depending
on the SUSY scale; compare Fig.\ref{HV1-energy-fig}b.
 Thus, the  $F_{---+}$  vanishing occurs earliest
for    "light SUSY";  later on for  SPS1a';  but it needs energies of
more than  10 TeV, in order to be seen for BBSSW.

The actual high energy behavior of the GBHV1 amplitudes is not given by
logarithmic expressions analogous to those in (\ref{ReFmmmm-LL}, \ref{ReFmpmp-LL}).
Nevertheless, asymptotic expressions
may be  obtained for them, by neglecting all masses in the
diagrammatic results  and using the asymptotic PV
functions of \cite{techpaper}.
Using these, we show in Figs.\ref{HV1-asym-fig}a,b, the GBHV1 1-loop
amplitudes for
 SM and the $SPS1a'$ MSSM model, and  compare them
to their asymptotic expressions denoted as "SM-asym" and
"$SPS1a'$-asym", respectively.

As seen in  Fig.\ref{HV1-asym-fig}a for  SM,
$F_{---+}$ remains almost constant in the whole
range $1\lsim \sqrt{s} \lsim 20 $TeV, with its  values almost
coinciding with those of the asymptotic expressions described above.
For $F_{---0}$ though, which presents an   $m_W/\sqrt{s}$ mass suppression
effect in the energy range of the figure,  there is a
considerable difference between the exact and
asymptotic expression. This is due to the mass suppressed
terms $m^2_W/s$, neglected in the asymptotic expression,
which, multiplied by the longitudinal helicity factor
$\sqrt{s}/m_W$, contributes  additional terms.

Correspondingly for $SPS1a'$, we see from Fig.6b that the
total MSSM amplitude $F_{---+}$ vanishes like $M^2/s$;
with $M$ describing some average SUSY scale for each benchmark.
This spectacular behavior is due to the cancellation
(expected from the HC rule \cite{heli}) between the
SM constant contribution and a similar, but opposite
SUSY constant contribution. In the case of $F_{---0}$,
the behavior is consistent with an $M/\sqrt{s}$ one.
For other benchmarks, the same features appear, using the
corresponding $M$ values. All these features can be
analyzed precisely using the explicit asymptotic forms
given in  \cite{techpaper}.

The angular distributions for the GBHV1 amplitudes
at $\sqrt{s}=0.5$TeV and $\sqrt{s}=4$TeV,
are shown in Figs.\ref{HV1-angle-fig}a and
\ref{HV1-angle-fig}b respectively.
The  shapes for SM and the  MSSM models are similar at
low energies, but rather different at high energies.
The SUSY cancellation mentioned above (spectacular for low
SUSY masses), considerably  reduces the backward peaking
at 4TeV. \\

\subsection{Features of the GBHV2 amplitudes $F_{-+--},F_{-+-0}$.}

The GBHV2 amplitudes $F_{-+--},~F_{-+-0}$  receive Born contributions,
which force them to vanish at high energy like
$\mw^2/s$ and $\mw/\sqrt{s}$ respectively, in agreement
with the HC rule; compare (\ref{F-Born-HV2-as}).

The energy dependence of the GBHV2 Born amplitudes,
as well as the 1-loop SM and MSSM amplitudes, at $\theta=60^o$,
are presented in Fig.\ref{HV2-energy-fig}a,b, for the same  energy ranges,
as before. Correspondingly, in Fig.\ref{HV2-asym-fig}a,b,
we compare the 1-loop and asymptotic values of the GBHV2 amplitudes in
SM and $SPS1a'$ respectively.

As seen from Fig.\ref{HV2-energy-fig}b,\ref{HV2-asym-fig}a,
the 1-loop SM result for  $F_{-+-0}$ is slowly vanishing like $1/\sqrt{s}$,
  above 3TeV;  while $F_{-+--}$ seems  to tend to a very small constant.

The  SUSY effects forcing the GBHV2 amplitudes to
vanish asymptotically, may be observed in Fig.\ref{HV2-energy-fig}b
and Fig.\ref{HV2-asym-fig}b.
In more detail,  Fig.\ref{HV2-energy-fig}b indicates that the tendency for the
MSSM $F_{-+--}$ amplitude   to vanish at high energies
is  obvious for the low scale models "light SUSY" and
$SPS1a'$; but for the "heavy scale" BBSSW,  higher energies are needed.
In contrast, the  $F_{-+-0}$ amplitude always vanishes like
$1/\sqrt{s}$, in all three MSSM benchmarks, as well as in SM.\\

The angular distributions for the GBHV2 amplitudes are shown in
Figs.\ref{HV2-angle-fig}a,b for the same
energy regimes as in Figs.\ref{HV1-angle-fig}a,b.
As seen in these figures, the GBHV2 amplitudes are notably different
from the GBHC amplitudes, as well as the GBHV1 ones.
The angular distribution is roughly model independent at 0.5TeV for both
$F_{-+--}$ and $F_{-+-0}$; but at 4 TeV, some model dependence appears
for $F_{-+--}$, whose strong backward dip  in SM, becomes
milder as the MSSM scale is reduced.

\vspace*{0.5cm}
\noindent
\underline{Overall conclusion for Section 3}.\\
We could claim that above (2-3)TeV,
the GBHC amplitudes completely dominate $ug\to dW^+$.
Moreover, the electroweak contribution to these amplitudes
is accurately described by (\ref{ReFmmmm-LL}, \ref{ReFmpmp-LL}), for both
SM and MSSM. The only model dependence in MSSM concerns
the value of $M_{SUSY}$.
At energies below 1 TeV though, the GBHV2 amplitudes  $F_{-+--}$ and
$F_{-+-0}$, may not be negligible.\\

\section{$W^{\pm}$ distributions at hadron colliders}

The relevant  subprocesses for $W^+$+jet
production induced by quarks of the
first family are\footnote{As already mentioned in Section 2,
we assume that the event sample does not include hard
photons emitted in association with $W$.}
\bq
ug\to dW^+ ~~,~~ \bar d g\to \bar u W^+~~,~~ \bar d u\to W^+ g~~,
\label{Wp-subprocesses}
\eq
while the conjugate subprocesses responsible  for $W^-$+jet production are
\bq
\bar u g \to \bar d W^- ~~, ~~  d g \to u W^- ~~,~~
 \bar u d \to g W^-~~. \label{Wm-subprocesses}
 \eq
One should then add the contributions of
the 2nd and 3rd families. Top quark processes
need  not be considered though, since  the  top PDF is negligible,
and a final top does not produce the same jets
as a light quark. In other words, a final top
can be   clearly separated and identified
as a different process \cite{singletop}. \\

Folding in the various  PDFs  and  the unpolarized
cross sections for the above subprocesses,
one can compute various types of
distributions (rapidities, angles, transverse momenta)
at a hadron collider. Here we concentrate on the
$W^{\pm}$  transverse momentum  distributions at the LHC,
in order  to show the size of the electroweak contribution to the SUSY effects,
as compared to the corresponding SM effects studied  in  \cite{kuhnW, HKK}.
This may be written as
\bq
{d\sigma(W^\pm+jet)\over dp_T}=\int^1_0 dx_a\int^1_0 dx_b
\theta(x_ax_b-\tau_m)~[P^{W^{\pm}}(x_a,x_b)
+\tilde P^{W^{\pm}}(x_a,x_b)] ~~, \label{LHC-pT}
\eq
where $p_T$ is the $W$ transverse momentum and
\bq
\tau_m={1\over S}\left (p_T+\sqrt{p^2_T+\mw^2}\right )^2 ~, \label{taum}
\eq
with $S$ being the total p-p energy squared at LHC, and
(s,t,u) being  the usual Mandelstam variable of
the subprocesses.

Denoting the LHC proton  PDFs for the various initial quarks,
antiquarks and gluons
as $f_q(x)$ etc,   the various subprocess
contributions to $W^{\pm}$ production in (\ref{LHC-pT}) may be written  as
\bqa
 P^{W^+}(x_a,x_b) &=&  {d\hat \sigma (u g\to d W^+)\over dp_T}
\Big [f_u(x_a)f_g(x_b)+f_c(x_a)f_g(x_b) \Big ] \nonumber \\
&  + & {d\hat\sigma (\bar d g\to \bar u W^+) \over dp_T}
\Big [f_{\bar d}(x_a)f_g(x_b)(|V_{ud}|^2+ |V_{cd}|^2)
\nonumber\\
& + & f_{\bar s}(x_a)f_g(x_b)(|V_{us}|^2+|V_{cs}|^2)
+f_{\bar b}(x_a)f_g(x_b)(|V_{ub}|^2+|V_{cb}|^2) \Big ]
\nonumber\\
&+ & {d\hat\sigma (u \bar d \to g W^+ )\over dp_T}
\Big [f_u(x_a)f_{\bar d}(x_b)|V_{ud}|^2+
f_c(x_a) f_{\bar d}(x_b)|V_{cd}|^2
\nonumber\\
& + &f_u(x_a) f_{\bar s}(x_b)|V_{us}|^2
+f_c(x_a) f_{\bar s}(x_b)|V_{cs}|^2
\nonumber\\
& + & f_u(x_a)f_{\bar b}(x_b)|V_{ub}|^2
+f_c(x_a) f_{\bar b}(x_b)|V_{cb}|^2 \Big ] ~, \label{PWp}\\
 P^{W^-}(x_a,x_b)& = &  {d\hat \sigma (u g\to d W^+)\over dp_T}
\Big [f_{\bar u}(x_a)f_g(x_b)+f_{\bar c}(x_a)f_g(x_b) \Big ]
 \nonumber\\
& +& {d\hat\sigma (\bar d g\to \bar u W^+) \over dp_T}
\Big [f_{ d}(x_a)f_g(x_b)(|V_{ud}|^2+ |V_{cd}|^2)
\nonumber\\
& + & f_{ s}(x_a)f_g(x_b)(|V_{us}|^2+|V_{cs}|^2)
+f_{ b}(x_a)f_g(x_b)(|V_{ub}|^2+|V_{cb}|^2) \Big ]
\nonumber\\
&+ &{d\hat\sigma (u \bar d \to g W^+ )\over dp_T}
\Big [f_{\bar u}(x_a)f_{ d}(x_b)|V_{ud}|^2+
f_{\bar c}(x_a) f_{ d}(x_b)|V_{cd}|^2
\nonumber\\
& + &f_{\bar u}(x_a) f_{ s}(x_b)|V_{us}|^2
+f_{\bar c}(x_a) f_{ s}(x_b)|V_{cs}|^2
\nonumber\\
& + & f_{\bar u}(x_a)f_{ b}(x_b)|V_{ub}|^2
+f_{\bar c}(x_a) f_{ b}(x_b)|V_{cb}|^2 \Big ] ~, \label{PWm}
\eqa
while
\bq
\tilde P^{W^{\pm}}(x_a,x_b)= P^{W^\pm}(x_b,x_a) ~~, \label{PWpm-tilde}
\eq
where  the $(x_a \leftrightarrow x_b)$ interchange
 only affects the arguments of the PDF's and
not the subprocess cross sections; compare (\ref{PWp}, \ref{PWm}),
and (\ref{RI}, \ref{RII}, \ref{RIII}) below. For the CKM matrix elements
$V_{mn}$, the unitarity relation  $\sum_n|V_{mn}|^2=1$ is also used.

The  contribution to (\ref{PWp}, \ref{PWm}) from the
subprocess  $ug \to dW^+$ is directly expressed in terms
of the helicity amplitudes in Section 2 as
\bqa
 {d\hat\sigma(ug\to dW^+)\over dp_T}&=& {p_T\over 768\pi s|t-u|}
\left [R_I|_{\theta}+R_I|_{\pi-\theta} \right ] ~, \nonumber  \\
 R_I(s,t,u)&= &\sum_{\lambda_u,\lambda_g,\lambda_d,\lambda_W}
|F_{\lambda_u\lambda_g\lambda_d\lambda_W}|^2 ~,~~\label{RI}
\eqa
where (\ref{kin-var}) is used together with the LHC kinematics
\bq
 s=Sx_ax_b ~~~,~~~ \cos\theta=\sqrt{1-\frac{4 p^2_T}{s\beta'^2}} ~~~,~~~
 |t-u|=s\beta'\sqrt{1-\frac{4 p^2_T}{s\beta'^2}}~~.~  \label{LHC-kin1}
\eq

It is important to note that the summation over  initial and final
helicity states in (\ref{RI}), guarantees that $R_I$
is an analytic function of $(s,t,u)$, with no
kinematical singularities related to the incoming or
outgoing nature of any particle state.
Thus, the usual  crossing
rules are applicable to $R_I$, which in turn allows the calculation
of all other subprocesses in (\ref{PWp}, \ref{PWm}).

The cross sections
for the other $W^+$ subprocesses are (compare (\ref{Wp-subprocesses}))
\bqa
 {d\hat\sigma(\bar d g \to \bar u W^+)\over dp_T}
&= & {p_T\over768\pi s|t-u|} \left [R_{II}|_{\theta}+R_{II}|_{\pi-\theta}\right ]
~~, \nonumber \\
 R_{II} &= & |R_I(u,  t, s)|~~, \label{RII}  \\
{d\hat\sigma(\bar d u \to  W^+g)\over dp_T} & = & {p_T\over 288\pi s|t-u|}
\left [R_{III}|_{\theta}+R_{III}|_{\pi-\theta} \right ] ~~, \nonumber \\
R_{III} &= & |R_I(t,  s, u)|~~. \label{RIII}
\eqa

Because of CP invariance, the corresponding cross sections
for the  $W^-$-production subprocesses are identical to
those for  $W^+$, but the PDFs obviously differ
for conjugate initial partons.

Just in order to show one example of SUSY effects,
we present in Fig.\ref{W-sig-fig}a ~
$d\sigma(pp\to W^\pm )/dp_T$ at  LHC,
in  the Born approximation, and  the 1-loop
SM and  $SPS1a'$ MSSM  model.
As is evident from this figure, the estimated production cross section
for either   $W^+$ or $W^-$, decreases as we move from the Born approximation,
to the 1-loop SM results and subsequently to  1-loop $SPS1a'$.

Actually, in all examples we have considered, the SUSY
contribution, reduces   the SM prediction.
In order to see this in more detail,
we plot in Fig.\ref{W-sig-fig}b
\bqa
D&= &{{d\sigma^{SM}(W^+)/dp_T-d\sigma^{MSSM}(W^+)/dp_T}
\over {d\sigma^{SM}(W^+)/dp_T}} \nonumber \\
& \simeq & {{d\sigma^{SM}(W^-)/dp_T-d\sigma^{MSSM}(W^-)/dp_T}
\over {d\sigma^{SM}(W^-)/dp_T}} ~, \label{ratio-D}
\eqa
versus $p_T$, for the three benchmark models of Table 1.

As seen from Fig.\ref{W-sig-fig}b, the \underline{reduction} $D$,
which is typically of the order of $10\%$, becomes stronger
as $M_{SUSY}$ gets smaller; \ie~ as we move from BBSSW, to $SPS1a'$
and then to the "light SUSY" model in Table 1. It is important to
emphasize, that in all cases, the  SUSY effect
 \underline{reduces} the SM expectation.

Moreover, Fig.\ref{W-sig-fig}b indicates that $D$ increases with $p_T$.
The observation of such a behavior though, does not necessarily points
towards SUSY for its origin, since similar effects may also
 be expected  in theories involving new gauge bosons or
extra dimensions \cite{beyond}. The identification of  a
SUSY effect could  only come after a detailed analysis
of many possible  observables; and of course, most importantly,
if SUSY sparticles are discovered at LHC.\\

One such observable may be  the
angular $W^\pm$ distribution in the subprocess c.m. system.
Restricting  for concreteness to  $W^+$ and in analogy to
(\ref{LHC-pT}-\ref{PWm}), this  is given by
\bq
{d\sigma(W^+ + jet)\over d s d\cos\theta }=\frac{1}{S} \int^1_{\frac{s }{S}}
\frac{dx_a}{x_a}
\left [ P^{W^+}_{ang}\left (x_a,\frac{s }{Sx_a},\theta \right ) +
\tilde P^{W^+}_{ang}\left (x_a,\frac{s }{Sx_a},\theta \right )\right ]
~~, \label{LHC-ang}
\eq
with
\bqa
P^{W^+}_{ang}(x_a,x_b,\theta) &=&
\Big [f_u(x_a)f_g(x_b)+f_c(x_a)f_g(x_b) \Big ]
{d\hat \sigma (u g\to d W^+)\over d\cos\theta } \nonumber \\
&  + &
\Big [f_{\bar d}(x_a)f_g(x_b)(|V_{ud}|^2+ |V_{cd}|^2)
 +  f_{\bar s}(x_a)f_g(x_b)(|V_{us}|^2+|V_{cs}|^2)
\nonumber \\
&+ & f_{\bar b}(x_a)f_g(x_b)(|V_{ub}|^2+|V_{cb}|^2) \Big ]
{d\hat\sigma (\bar d g\to \bar u W^+) \over d\cos\theta }
\nonumber\\
&+ & \Big [f_u(x_a)f_{\bar d}(x_b)|V_{ud}|^2+
f_c(x_a) f_{\bar d}(x_b)|V_{cd}|^2
\nonumber\\
& + &f_u(x_a) f_{\bar s}(x_b)|V_{us}|^2
+f_c(x_a) f_{\bar s}(x_b)|V_{cs}|^2
\nonumber\\
& + & f_u(x_a)f_{\bar b}(x_b)|V_{ub}|^2
+f_c(x_a) f_{\bar b}(x_b)|V_{cb}|^2 \Big ]
{d\hat\sigma (u \bar d \to g W^+ )\over d\cos\theta} ~, \label{PWp-ang}
\eqa
\bq
\tilde P^{W^+}_{ang}(x_a,x_b,\theta )= P^{W^+}_{ang}
(x_b,x_a, \pi-\theta) ~~, \label{PWp-tilde}
\eq
\bqa
&&  {d\hat\sigma(u g\to d W^+) \over \cos\theta } =
\frac{\beta'}{3072\pi s}  [R_I|_{\theta}]  ~, \nonumber \\
&&  {d\hat\sigma(\bar d g\to \bar u W^+) \over \cos\theta } =
\frac{\beta'}{3072\pi s}  [R_{II}|_{\theta}]  ~, \nonumber \\
&&  {d\hat\sigma(u \bar d \to g W^+) \over \cos\theta } =
\frac{\beta'}{1152\pi s} [R_{III}|_{\theta}]    ~ . \label{dsig-ang}
\eqa
As in the (\ref{PWp},\ref{PWm})-cases,
the corresponding  distribution for  $W^-$ is  obtained from
(\ref{PWp-ang}) by changing each parton distribution function to
the corresponding anti-parton. And the percentage decrease
of the SM result induced by SUSY is given, in analogy to (\ref{ratio-D}), by
\bqa
D_{ang}&= &{{d\sigma^{SM}(W^+)/ds d\cos\theta
-d\sigma^{MSSM}(W^+)/ds d\cos\theta}
\over {d\sigma^{SM}(W^+)/ds d\cos\theta}} \nonumber \\
& \simeq  & {{d\sigma^{SM}(W^-)/ds d\cos\theta-d\sigma^{MSSM}(W^-)/ds d\cos\theta}
\over {d\sigma^{SM}(W^-)/ds d\cos\theta}} ~. \label{ratio-Dang}
\eqa
The corresponding results are shown in Figs.\ref{W-angsig-fig}a,b
for the angular distribution and the percentage reduction $D_{ang}$
at a subprocess c.m. energy of 0.5TeV; while the results
at Figs.\ref{W-angsig-fig}c,d  apply to a  subprocess
c.m. energy of  4 TeV. As seen there,
if $M_{SUSY}$ is not too high, the SUSY reduction of the SM
prediction is at the 10\% level. It increases with the
subprocess energy especially in the central region
($\theta\simeq90^o$).   \\

In an actual $W$ production experiment, we should also include
the infrared QED, QCD and higher order effects, like those
partially calculated by \cite{kuhnW, HKK}. These effects are
to a large extent detector dependent, and have to be considered
in conjunction with the specific experiment carried.
In any case this should not affect the properties and the
size of the SUSY effects considered in the present paper.
A $10\%$ effect  should be largely visible, since it is
much larger than the statistical errors one gets from
the size of the cross section  given in Fig.\ref{W-sig-fig}a and an
integrated luminosity of 10 or 100 $fb^{-1}/{\rm year}$, expected
at LHC. Correspondingly for the angular distribution effects,
particularly for the lower energy region presented in
Fig.\ref{W-angsig-fig}a,b. \\

As already said,  more observables, like rapidities,
angles and ($W$+jet)-mass distributions
should also be considered in a detail experimental analysis.
The  angular distributions in particular, may be helpful  in
discriminating between  the GBHC and  GBHV2  amplitudes $F_{-+--}$,
$F_{-+-0}$, which are not  negligible, for energies below 1 TeV.
On the other hand, GBHV1 amplitudes seem to be negligible,
in the whole LHC range. In any case, an experimental measurement
of the angular distribution should confirm the dominance of
the GBHC amplitudes and the absence of any  anomalous
GBHV contribution.\\

\section{Conclusions and outlook}

In this paper we have underlined several
remarkable features of the process $ug\to dW^+$,
 at the tree and 1-loop electroweak level.\\

At Born level,
only the two GBHC amplitudes  $(F_{----},~ F_{-+-+})$ survive at high energy,
in agreement with the HC rule.
On the contrary, at the same Born level,
$(F_{---+},~ F_{---0})$
vanish  identically, while the remaining amplitudes are mass-suppressed as
\[
F_{-+--} \sim \frac{m^2_W}{s}~~ ,~~  F_{-+-0} \sim \frac{m_W}{\sqrt{s}}~~.
\]\\

At the  1-loop level in SM, the electroweak corrections modify
the two  GBHC amplitudes at high energy,
in accordance with  the logarithmic rules;
compare (\ref{ReFmmmm-LL},\ref{ReFmpmp-LL}) and Fig.\ref{HC-asym-fig}a.
These imply   corresponding reductions of the  GBHC amplitudes.

As far as the GBHV amplitudes in SM are concerned,
 ($F_{---+}$, $F_{-+--}$) behave like constants
 at high energy; while ($F_{---0}$, $F_{-+-0}$), which
   involve a longitudinal  $W$, vanish like $m_W/\sqrt{s}$;
see Figs.\ref{HV1-asym-fig}a, \ref{HV2-asym-fig}a.\\

The 1-loop SUSY contribution, at low energy, induces a bigger or smaller
reduction to  the SM amplitudes, depending
on the scale $M_{SUSY}$.

\begin{figure}[t]
\vspace*{-1.cm}
\[
\hspace{-0.5cm}
\epsfig{file=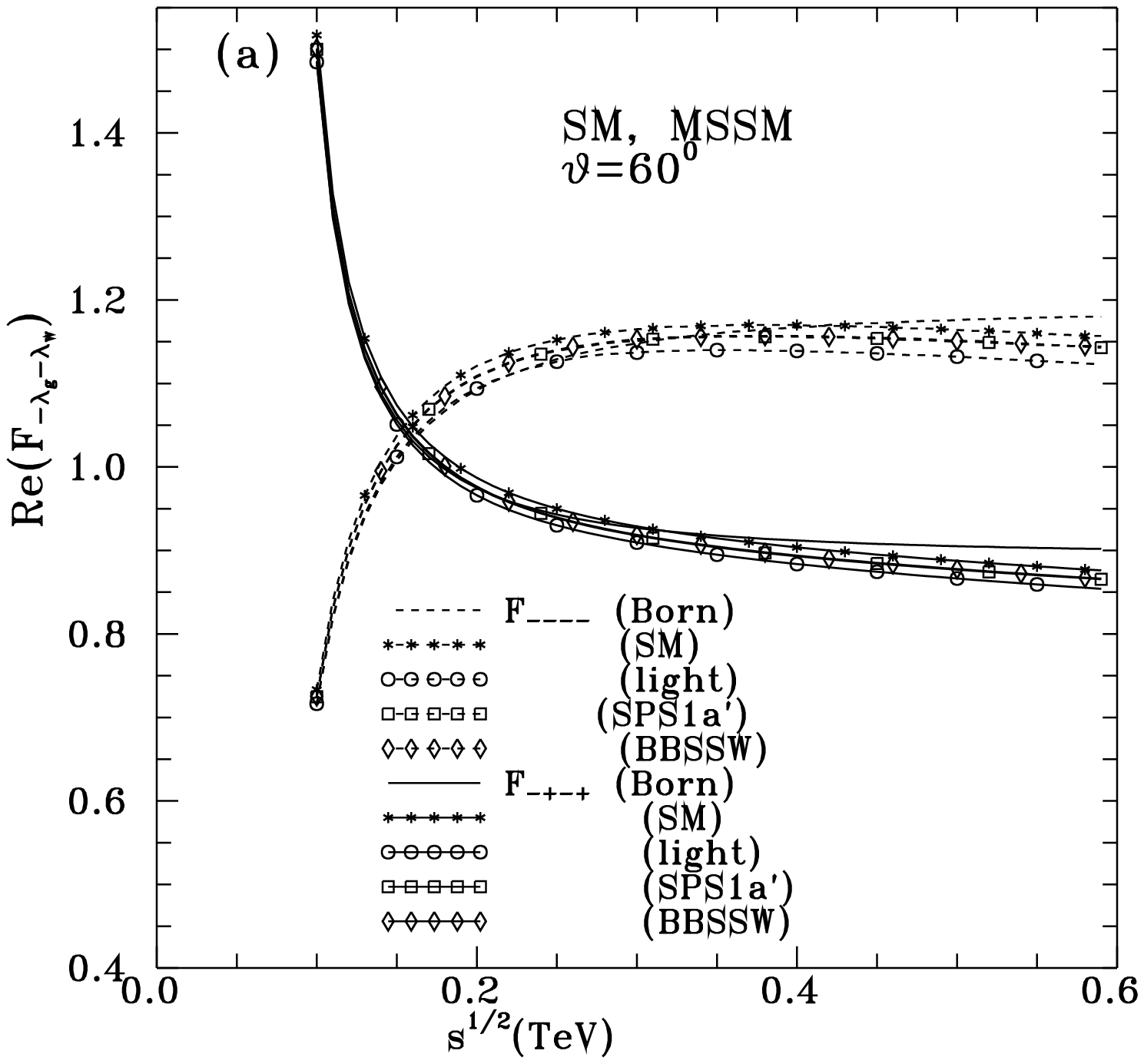,height=7.5cm}\hspace{0.5cm}
\epsfig{file=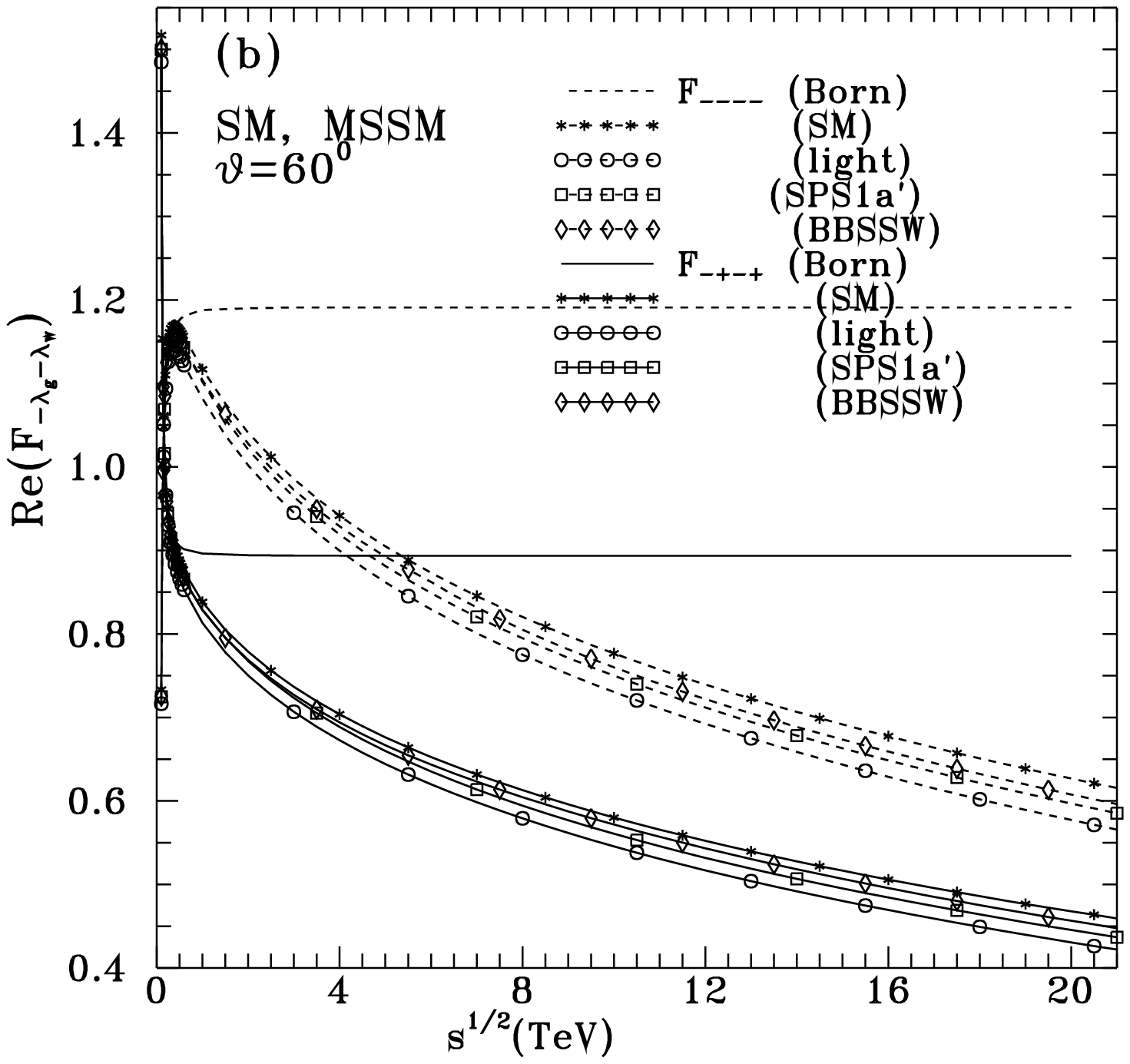,height=7.5cm}
\]
\caption[1]{Energy dependence of the helicity
conserving GBHC $ug\to dW$ amplitudes in the Born approximation
 and  the  1-loop  SM and   MSSM benchmark predictions.
 The c.m scattering angle is chosen
at $\theta=60^o$, while (a) and (b) cover respectively
 the LHC and the beyond LHC energy ranges. The coefficient
 $\lambda^a/2$ has been factored out in this figure
and in all the following ones.}
\label{HC-energy-fig}
\end{figure}

\begin{figure}[b]
\[
\hspace{-0.5cm}
\epsfig{file=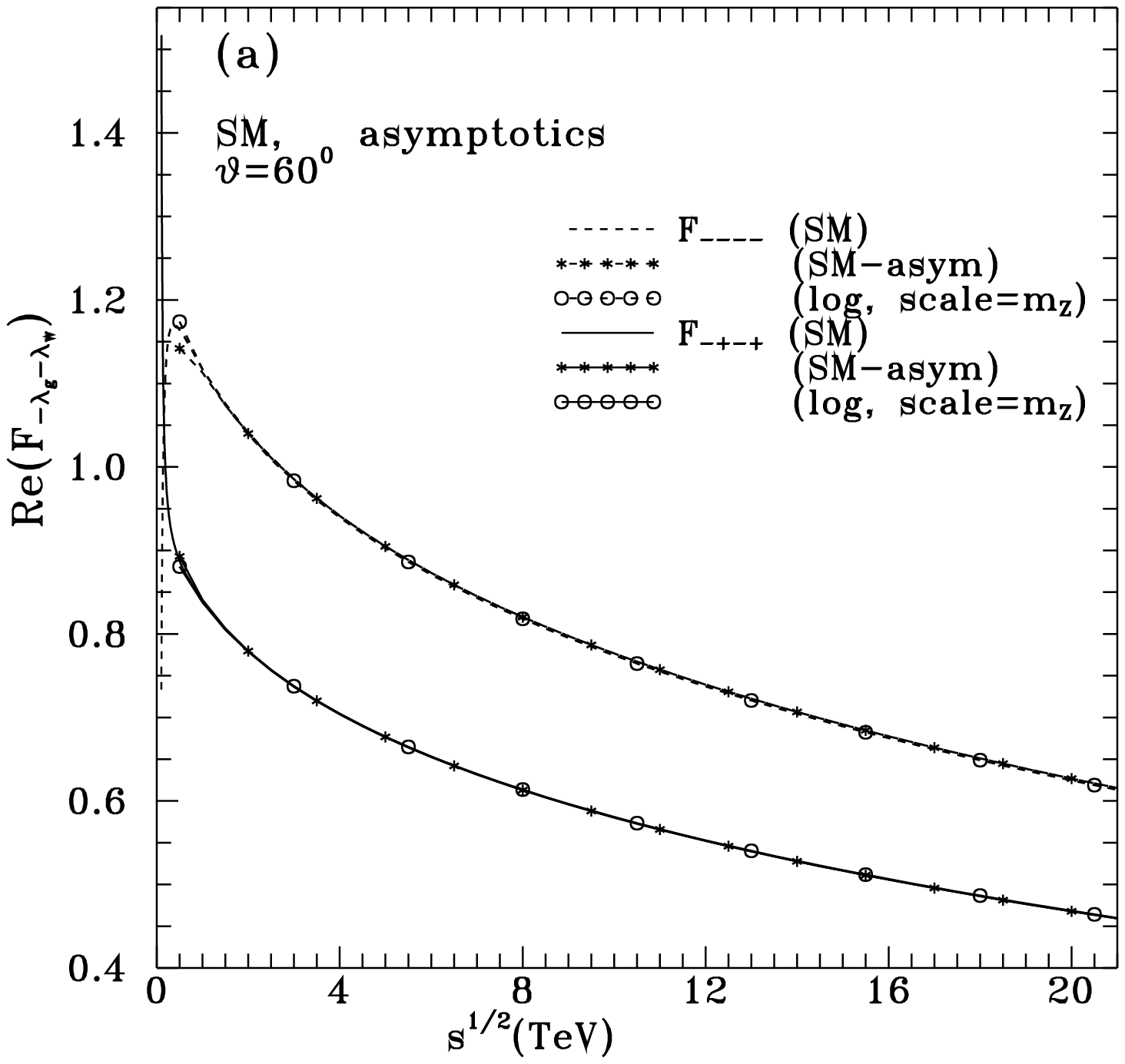,height=7.5cm}\hspace{0.5cm}
\epsfig{file=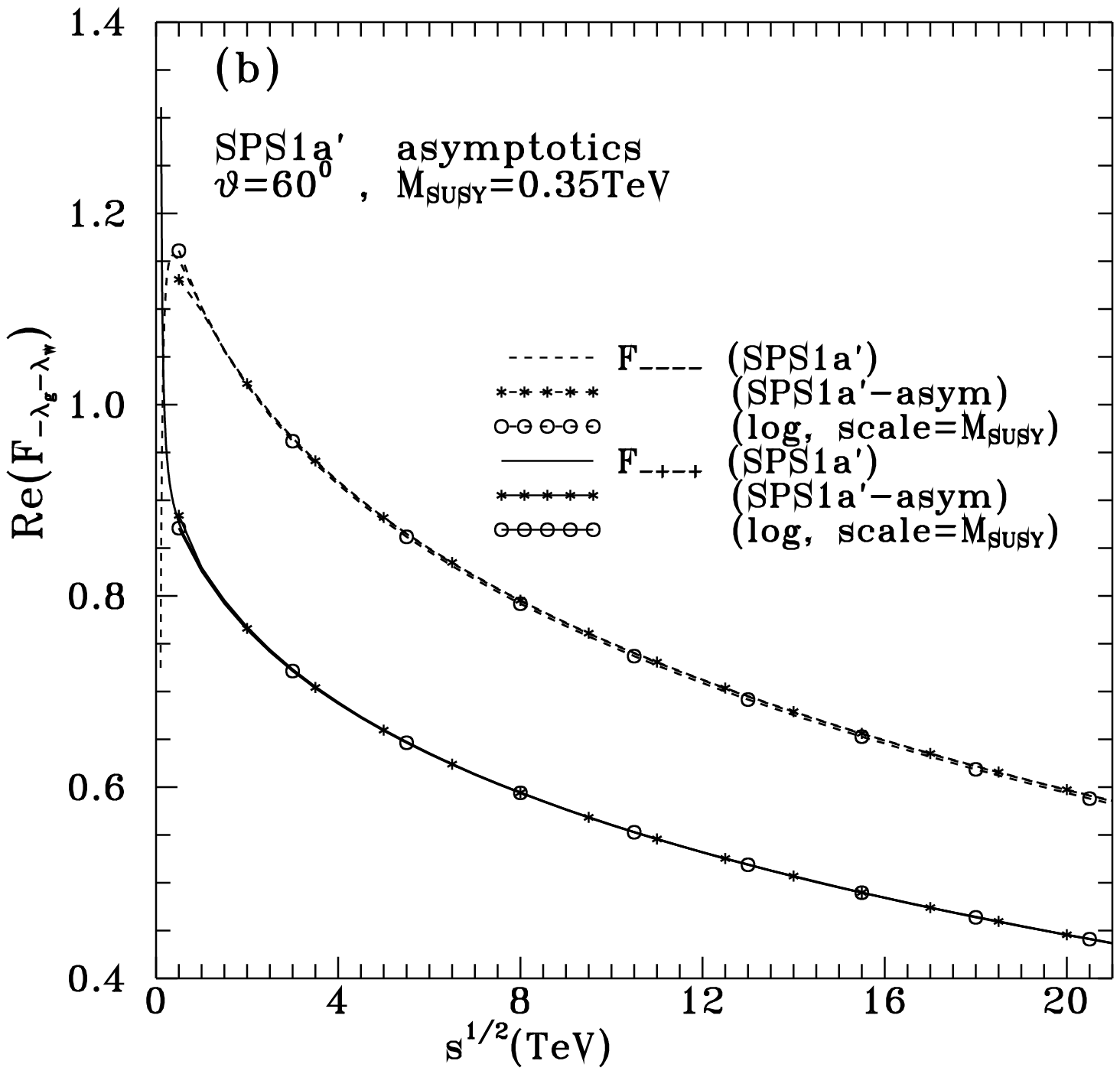,height=7.5cm}
\]
\caption[1]{High energy
dependence of the GBHC  $ug\to dW$ amplitudes at
$\theta=60^o$, in 1-loop SM (a), and an  MSSM benchmark model (b),
together with the corresponding leading log predictions using
$M_{SUSY}$ given in Table 1 and the "constants" in Table 2.}
\label{HC-asym-fig}
\end{figure}

\begin{figure}[t]
\vspace*{-1.cm}
\[
\hspace{-0.5cm} \epsfig{file=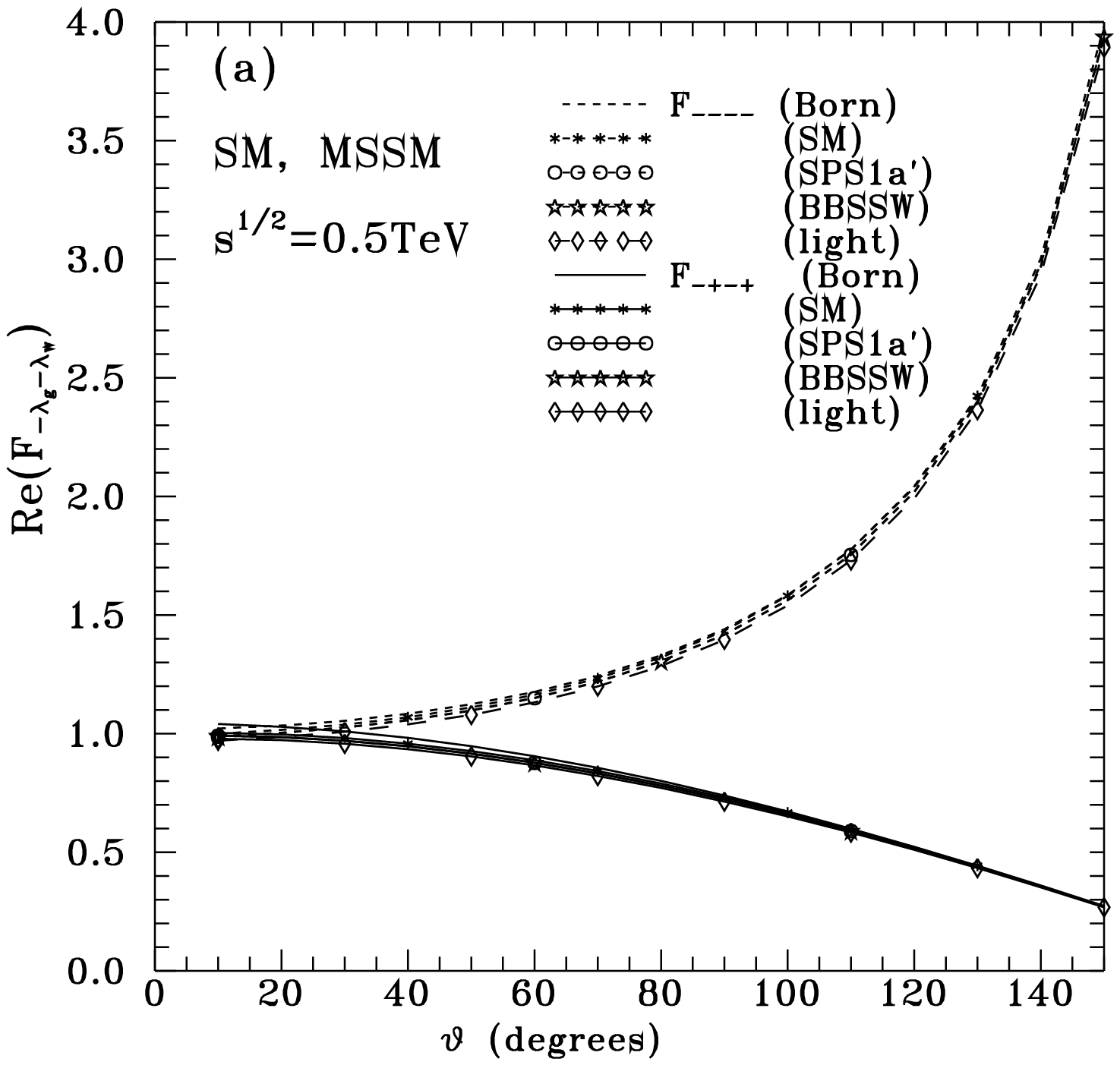,height=7.5cm}\hspace{0.5cm}
\epsfig{file=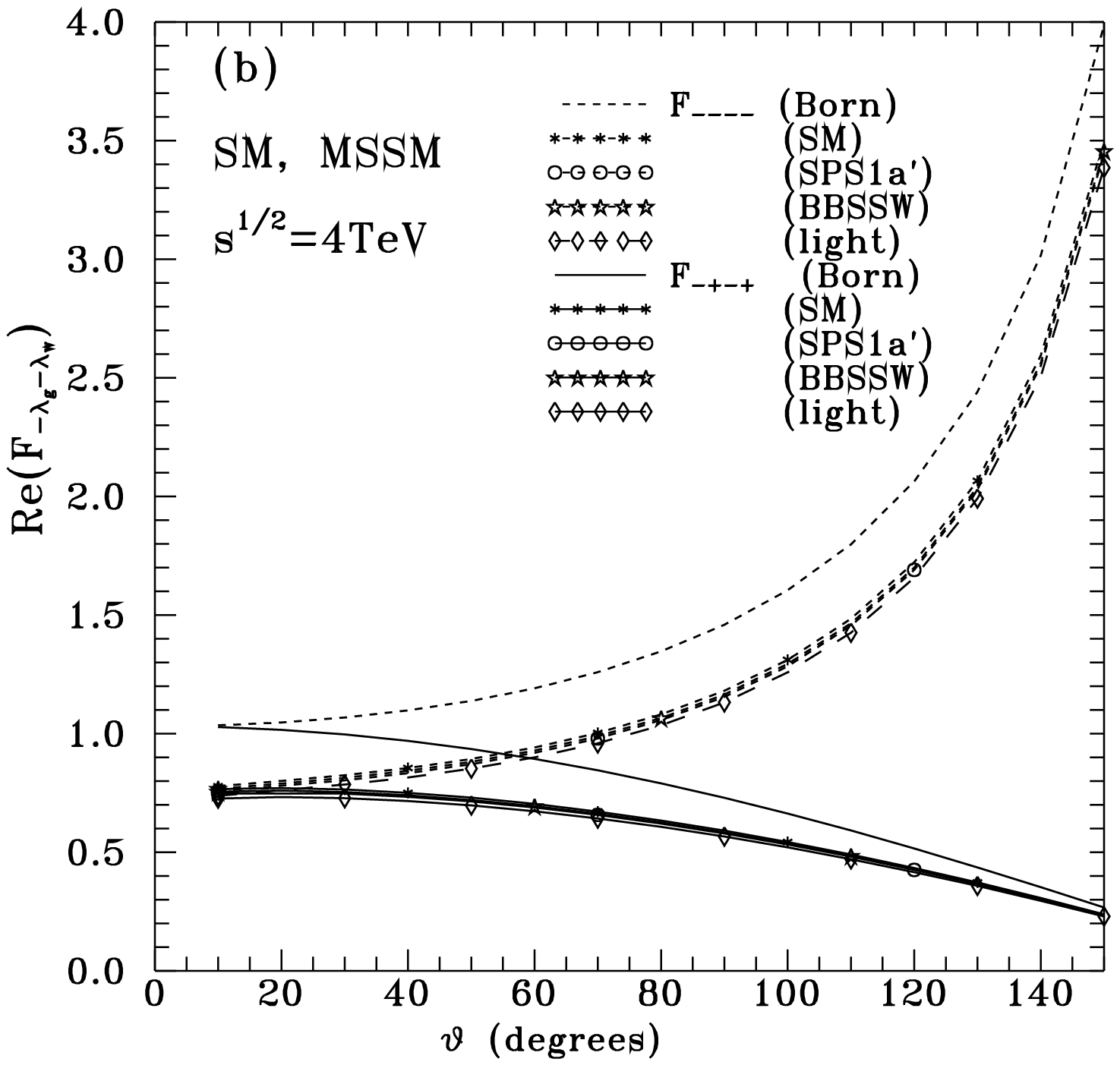,height=7.5cm}
\]
\caption[1]{Angular  dependence
of the GBHC  $ug\to dW$ amplitudes
in the Born approximation,  SM and   MSSM benchmark models, at
  c.m. energies   0.5 TeV (a), and 4 TeV (b).}
\label{HC-angle-fig}
\end{figure}

At energies comparable to $M_{SUSY}$ though,
remarkable features appear.
For the leading GBHC amplitudes,  negative
SUSY contributions arise, which
grow typically like $-\ln(s/M^2_{SUSY})$, in agreement
with the general SUSY asymptotic rules; compare
(\ref{ReFmmmm-LL},\ref{ReFmpmp-LL}) and
Figs.\ref{HC-energy-fig}b, \ref{HC-asym-fig}b.

As far as  the  transverse
GBHV amplitudes ($F_{---+}$, $F_{-+--}$) are concerned,
the SUSY contributions tend   asymptotically to constants,
which are  exactly opposite to the SM asymptotic constants.
As a result, the transverse  GBHV amplitudes   are
mass suppressed in  MSSM, like $M^2_{SUSY}/s$.
An analogous behavior is valid for the longitudinal GBHV amplitudes
($F_{---0}$, $F_{-+-0}$),
for which the   SUSY contributions are  also mass suppressed like
$M_{SUSY}/\sqrt{s}$, so that  the complete MSSM contribution again tends to zero.\\

These  remarkable
features constitute   a new illustration of the general HC rule
established in \cite{heli}. It  clearly indicates that even
in the presence of masses and electroweak gauge symmetry breaking,
the HC theorem remains correct; \ie~
all  two-body amplitudes violating the conservation of the
total helicity
 should vanish in  MSSM, and tend to constants  in SM.
A similar behavior has also been observed
in the complete 1-loop treatment of $\gamma \gamma \to ZZ$
and $\gamma \gamma \to \gamma Z$  \cite{heli, ggVV}.
In other words,  ratio-of-mass terms, which could be imagined to  spoil
the exact validity of  HC theorem in MSSM, are not  generated. This may be related
to the fact that physical amplitudes do not possess mass
singularities.

Contrary to  other conservation properties in particle physics,
which are  related to the existence of a continuous symmetry transformation
and  derived through Noether's construction for any physical processes;
HC is intimately related to 2-to-2 body processes induced by any 4-dimensional
softly broken supersymmetric extension of the standard model.
The validity of HC is not derived on the basis of the Lagrangian
of the model, but rather comes from an  analysis of all contributing diagrams,
to any order in perturbation theory.\\

\begin{figure}[b]
\[
\hspace{-0.5cm}
\epsfig{file=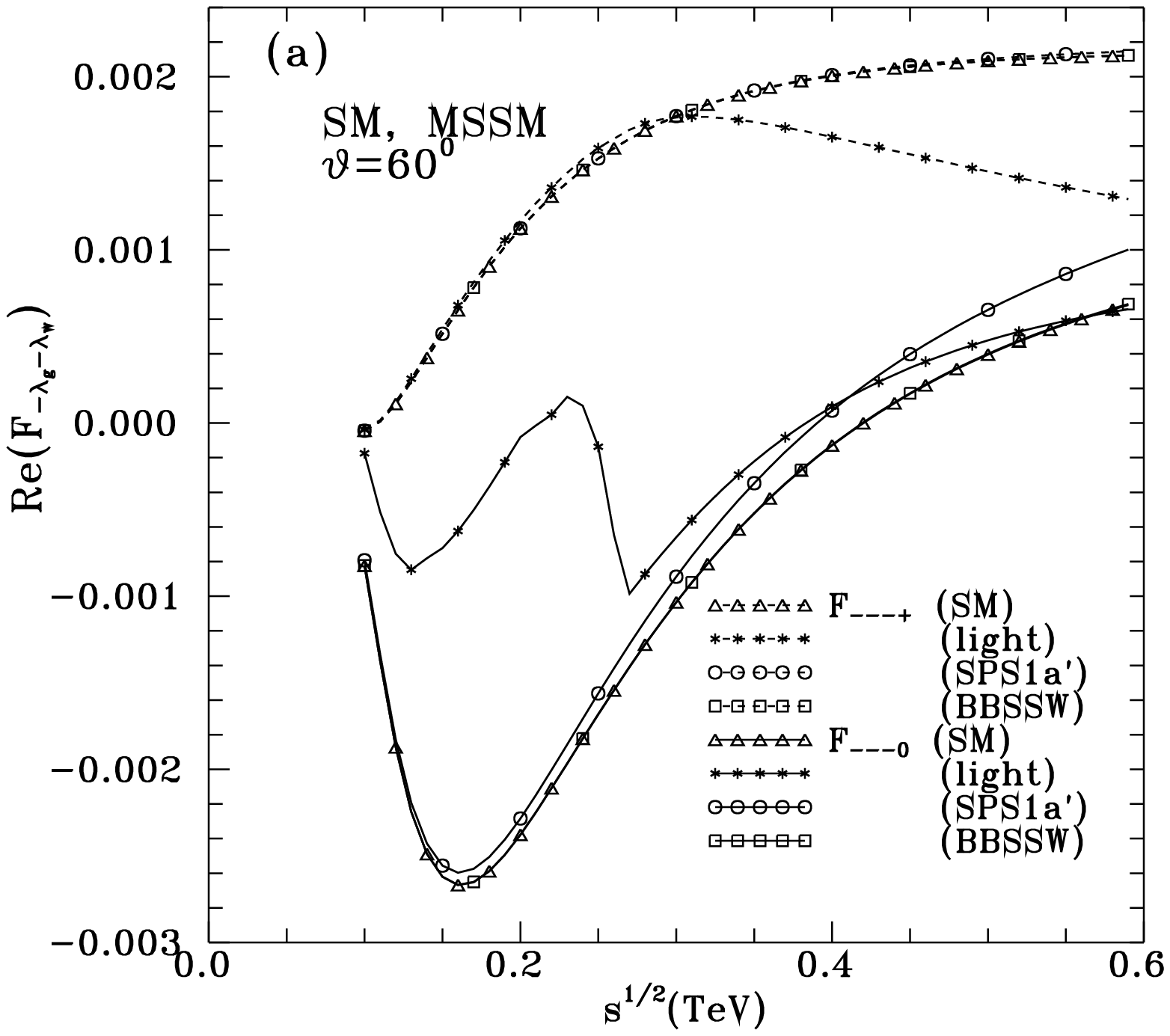,height=7.cm}\hspace{0.5cm}
\epsfig{file=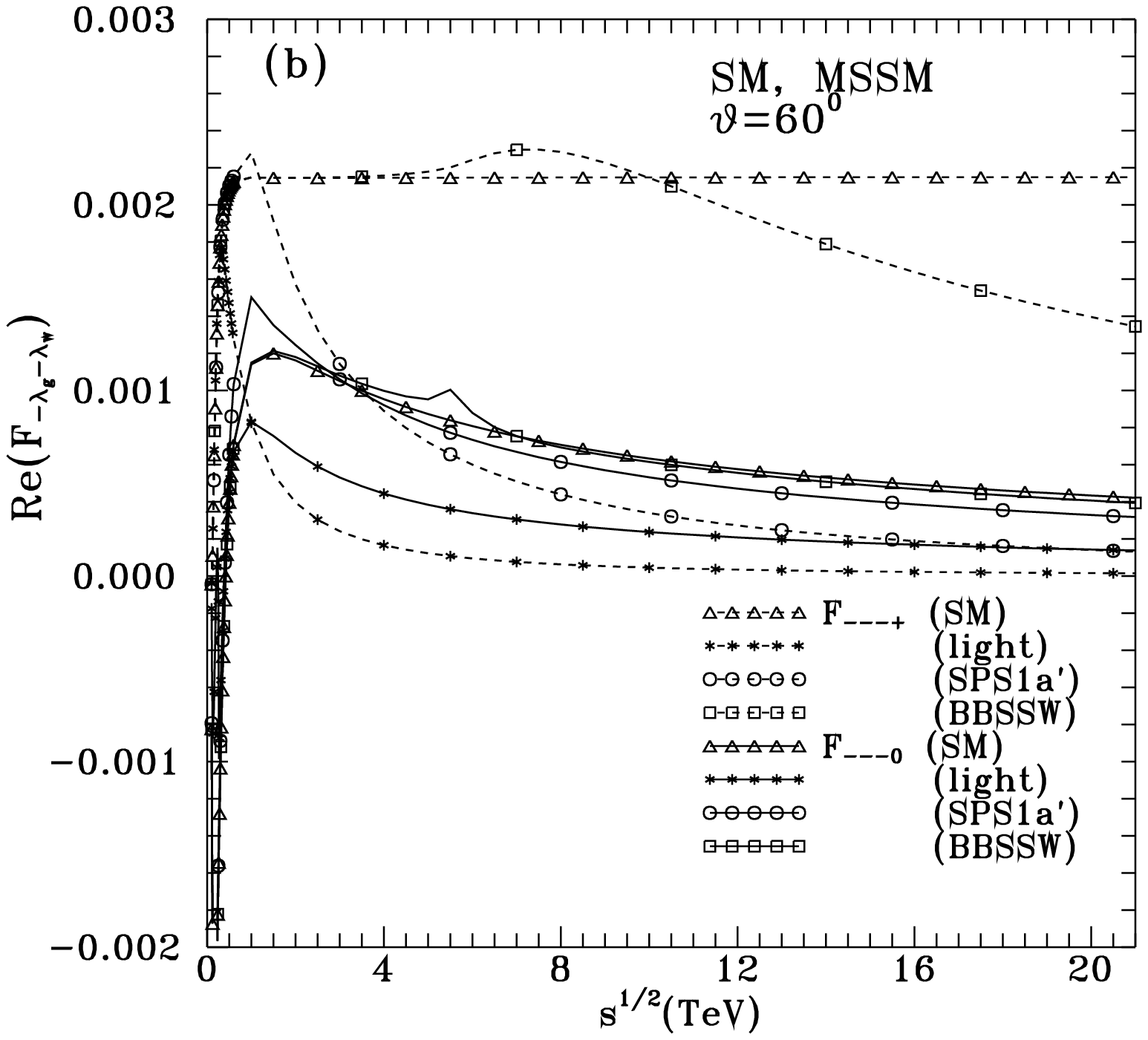,height=7.cm}
\]
\caption[1]{Energy dependence of the GBHV1 $ug\to dW$ amplitudes
$F_{---+},~F_{---0}$ at $\theta=60^o$,
in 1-loop SM and  3 MSSM benchmark models,
 for c.m. energies up to  0.6 TeV (a), and up to  21 TeV (b). }
\label{HV1-energy-fig}
\end{figure}

\begin{figure}[t]
\vspace*{-1.cm}
\[
\hspace{-0.5cm}
\epsfig{file=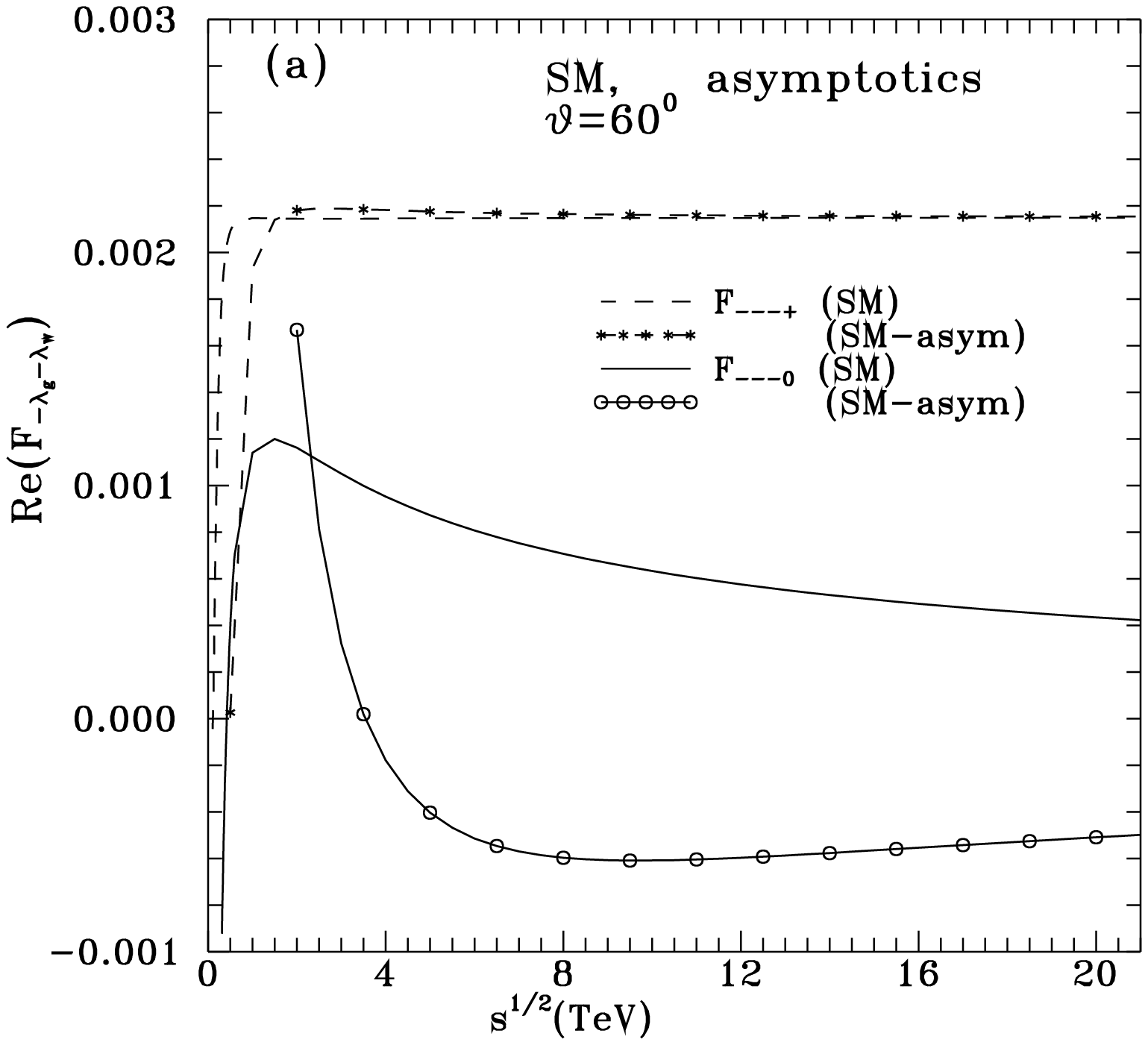,height=7.cm}\hspace{0.5cm}
\epsfig{file=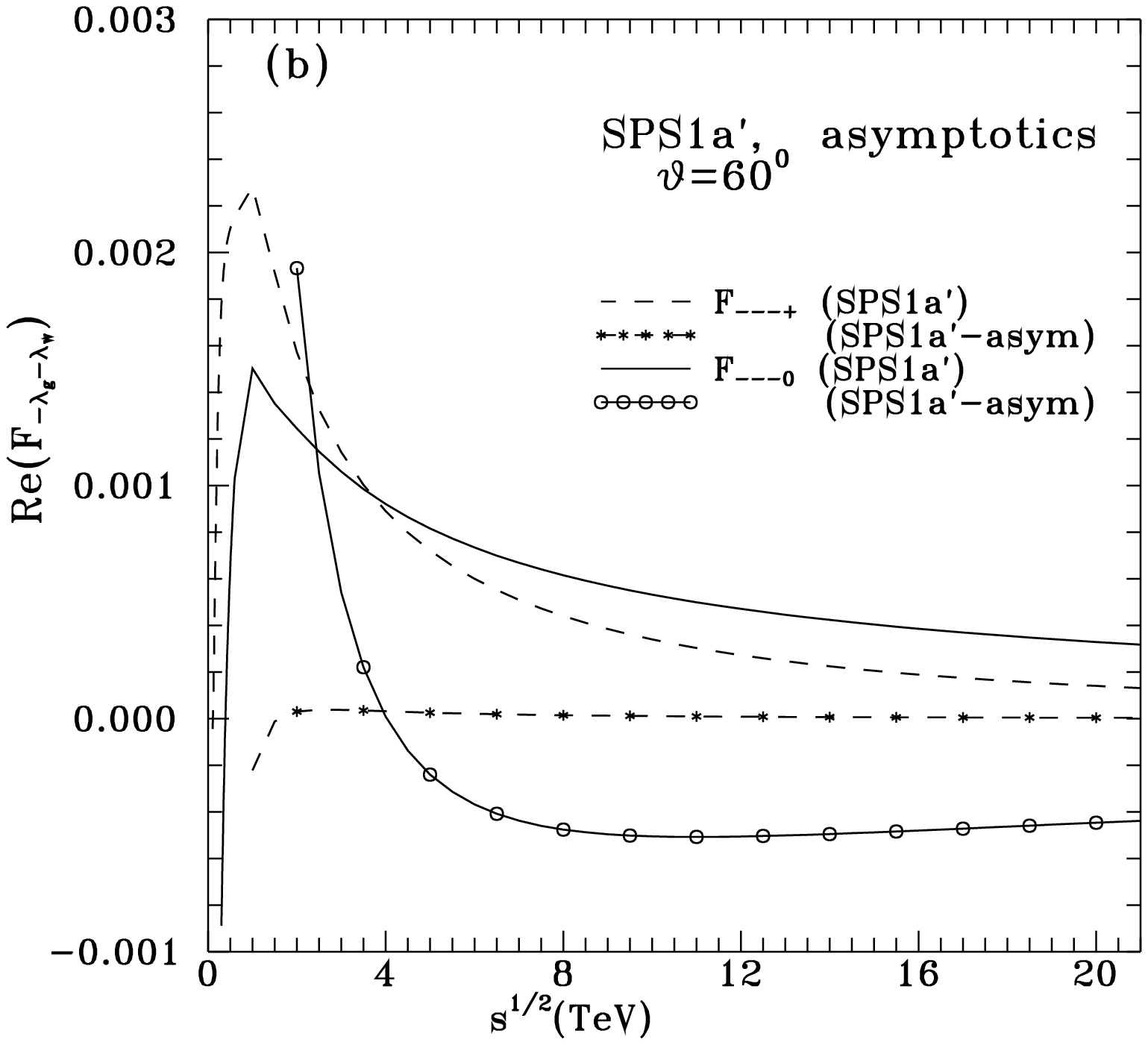,height=7.cm}
\]
\caption[1]{High energy
dependence of the helicity violating GBHV1  amplitudes $F_{---+},~F_{---0}$ at
$\theta=60^o$, in 1-loop SM (a), and $SPS1a'$   MSSM model (b).
The exact 1-loop results are compared to
the asymptotic ones   described in the text.}
\label{HV1-asym-fig}
\end{figure}

\begin{figure}[b]
\[
\hspace{-0.5cm} \epsfig{file=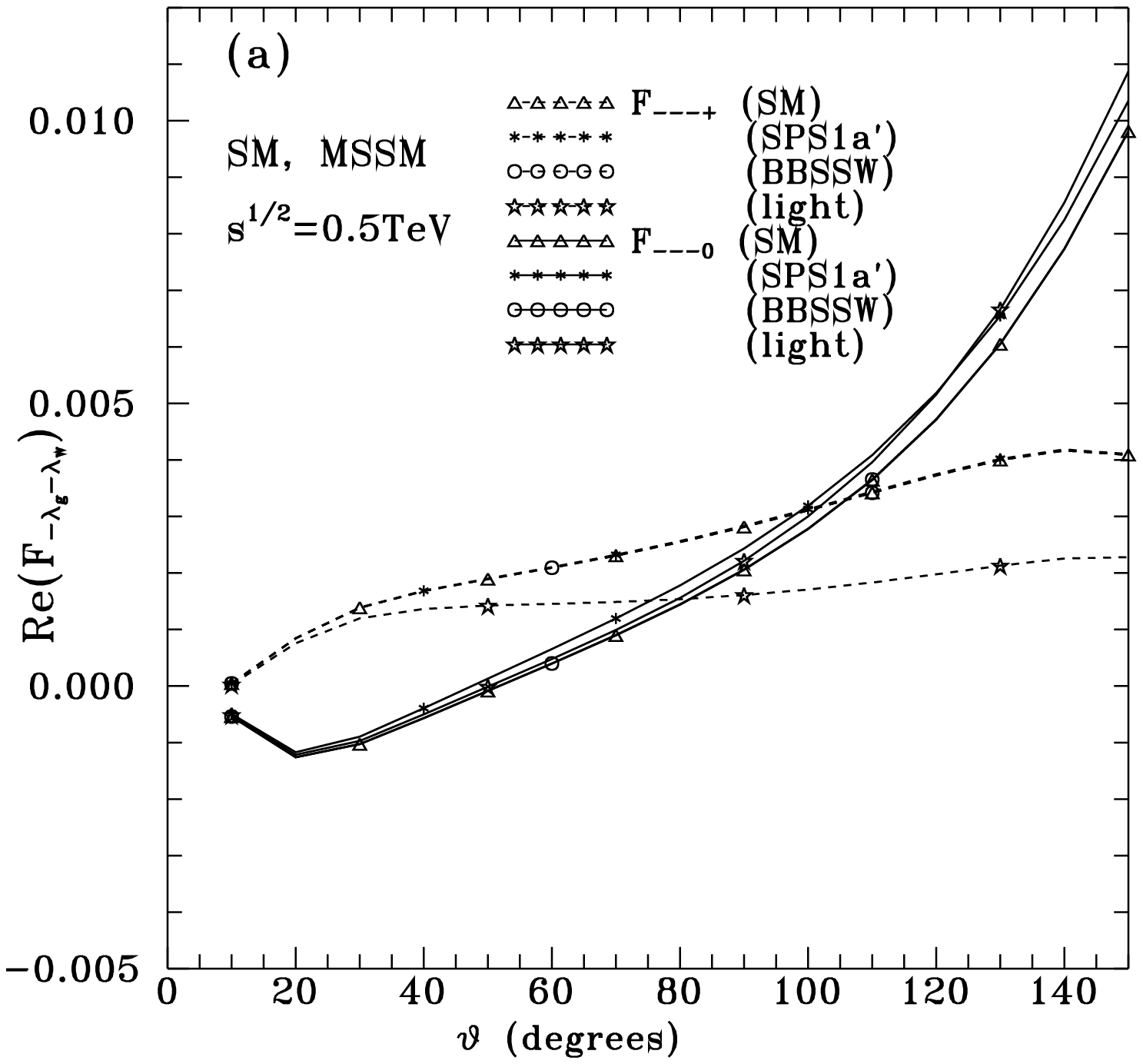,height=7.5cm}\hspace{0.5cm}
\epsfig{file=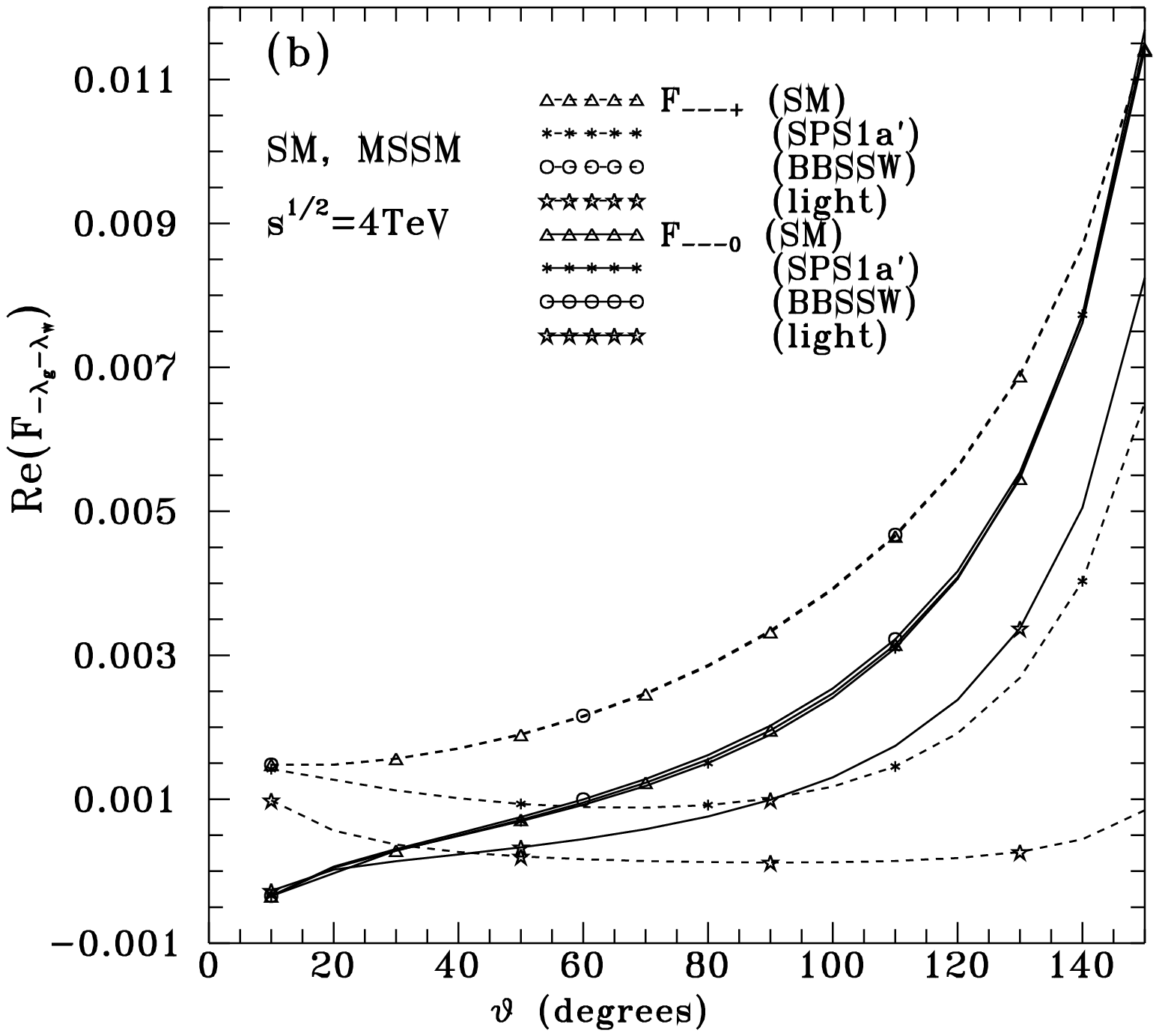,height=7.5cm}
\]
\caption[1]{Angular  dependence
of the helicity violating GBHV1 amplitudes $F_{---+},~F_{---0}$
for 1-loop SM and  three  MSSM benchmark models, at
  c.m. energies   0.5 TeV (a), and 4 TeV (b).}
\label{HV1-angle-fig}
\end{figure}

\begin{figure}[t]
\vspace*{-1.cm}
\[
\hspace{-0.5cm}
\epsfig{file=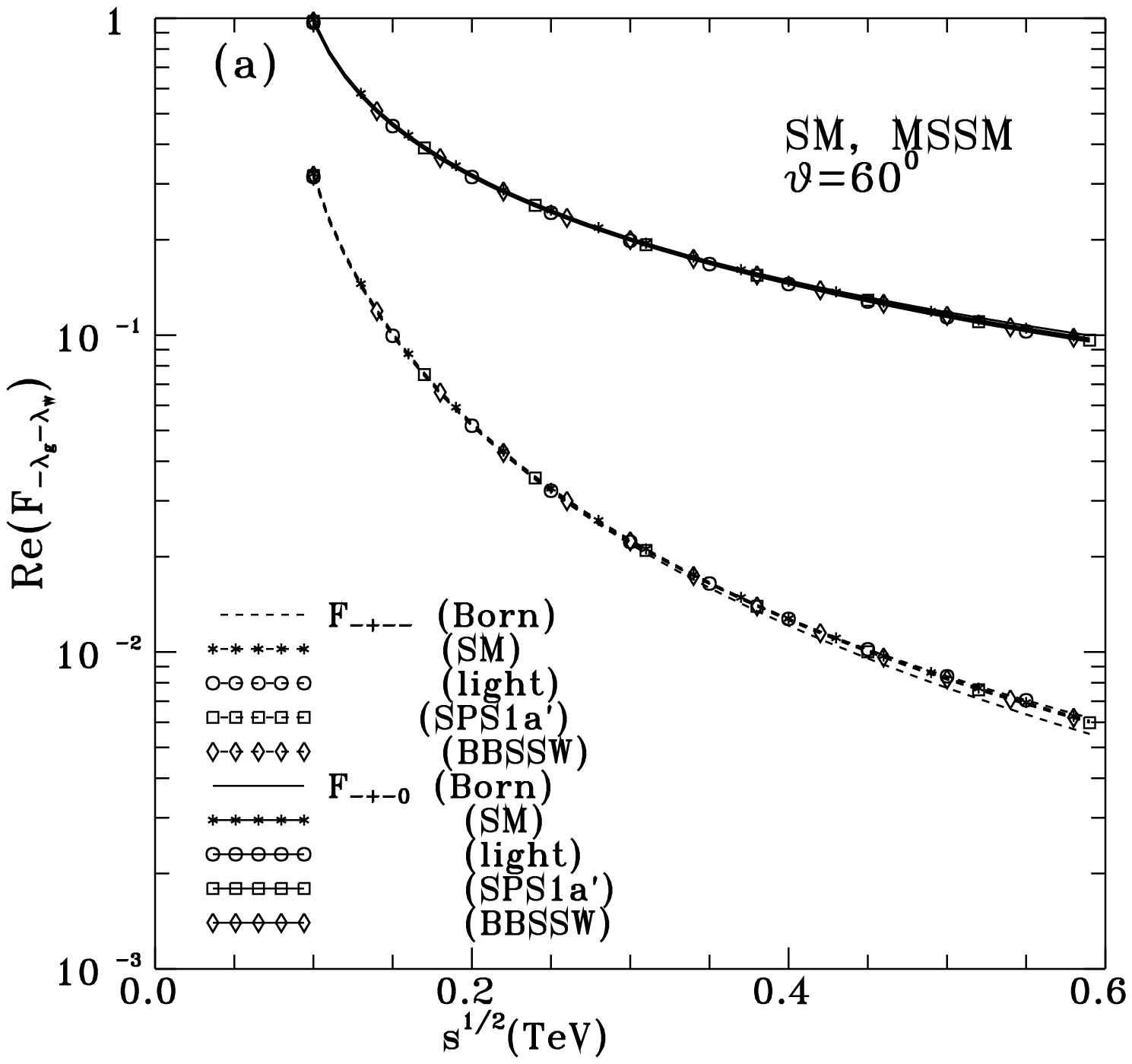,height=7.5cm}\hspace{0.5cm}
\epsfig{file=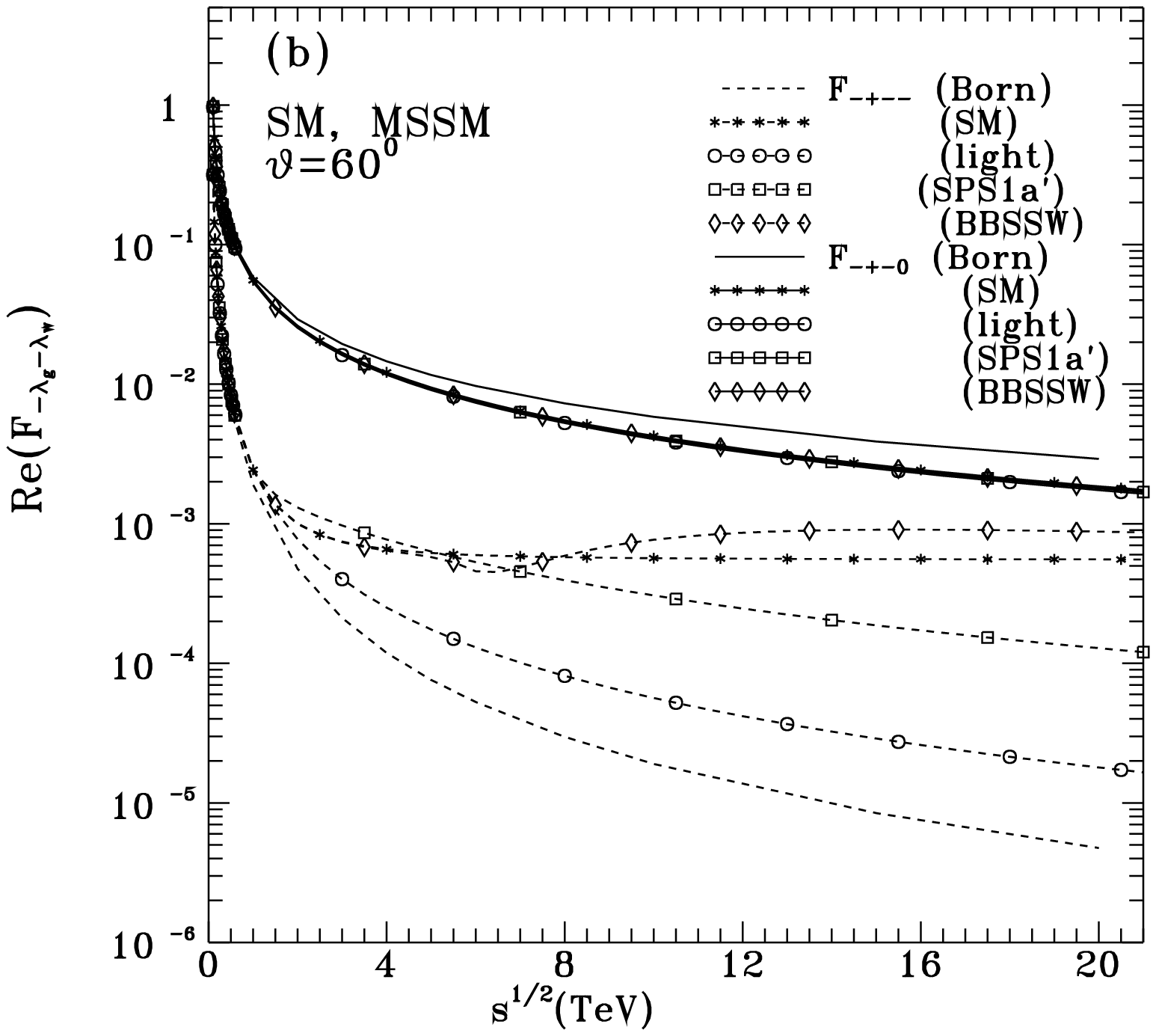,height=7.5cm}
\]
\caption[1]{Energy dependence of the helicity
violating GBHV2 $ug\to dW$ amplitudes
$F_{-+--},~F_{-+-0}$ at
$\theta=60^o$, in the Born approximation
and  1-loop SM and   MSSM benchmark models,
 for c.m. energies up to  0.6 TeV (a), and up to  21 TeV (b). }
\label{HV2-energy-fig}
\end{figure}

\begin{figure}[b]
\[
\hspace{-0.5cm}
\epsfig{file=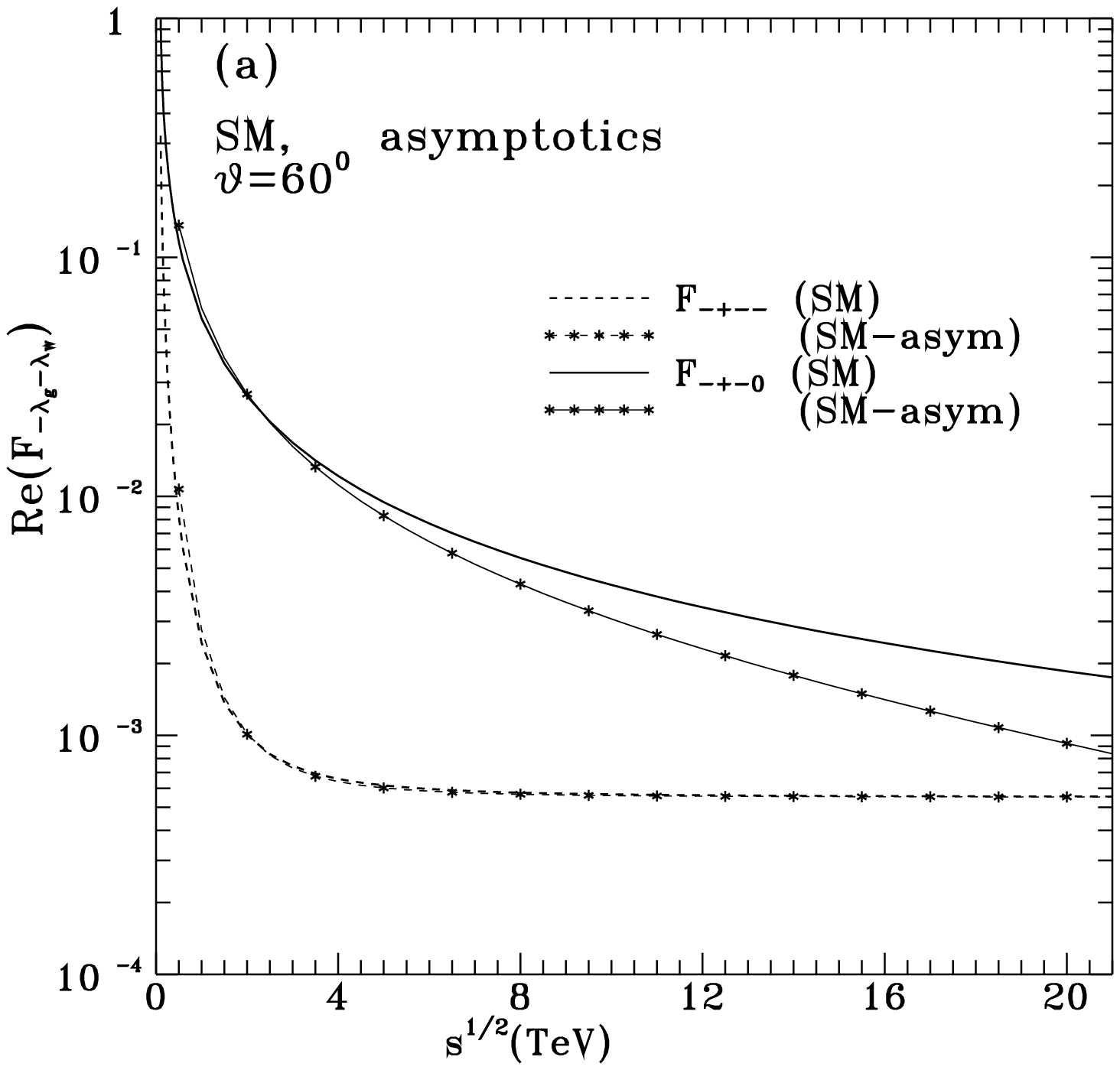,height=7.5cm}\hspace{0.5cm}
\epsfig{file=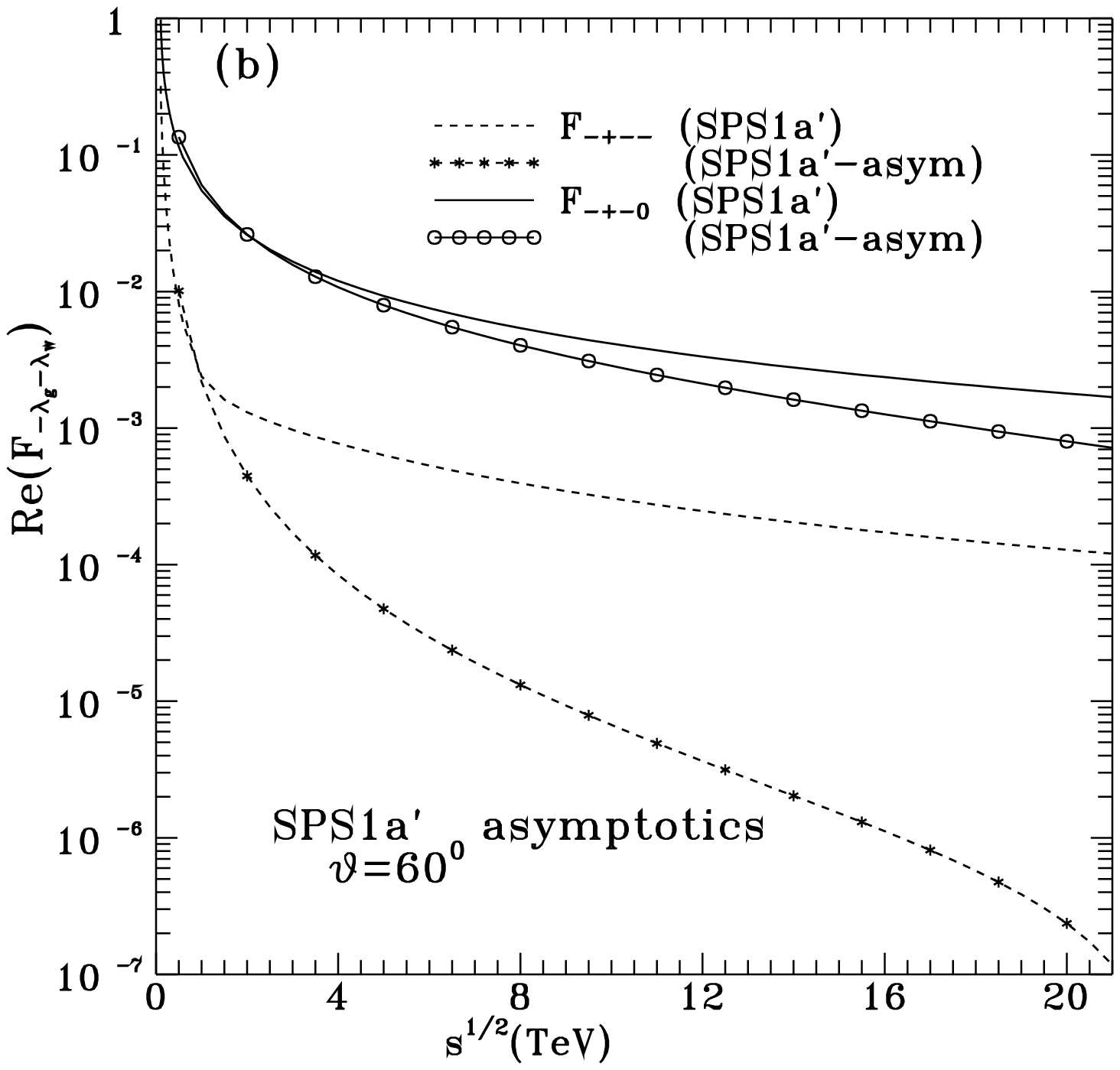,height=7.5cm}
\]
\caption[1]{High energy
dependence of the helicity violating GBHV2 amplitudes $F_{-+--},~F_{-+-0}$ at
$\theta=60^o$, in SM (a), and the $SPS1a'$   MSSM  model (b).
The exact 1-loop results are compared to
the asymptotic ones   described in the text. }
\label{HV2-asym-fig}
\end{figure}

\begin{figure}[t]
\vspace*{-1.5cm}
\[
\hspace{-0.5cm} \epsfig{file=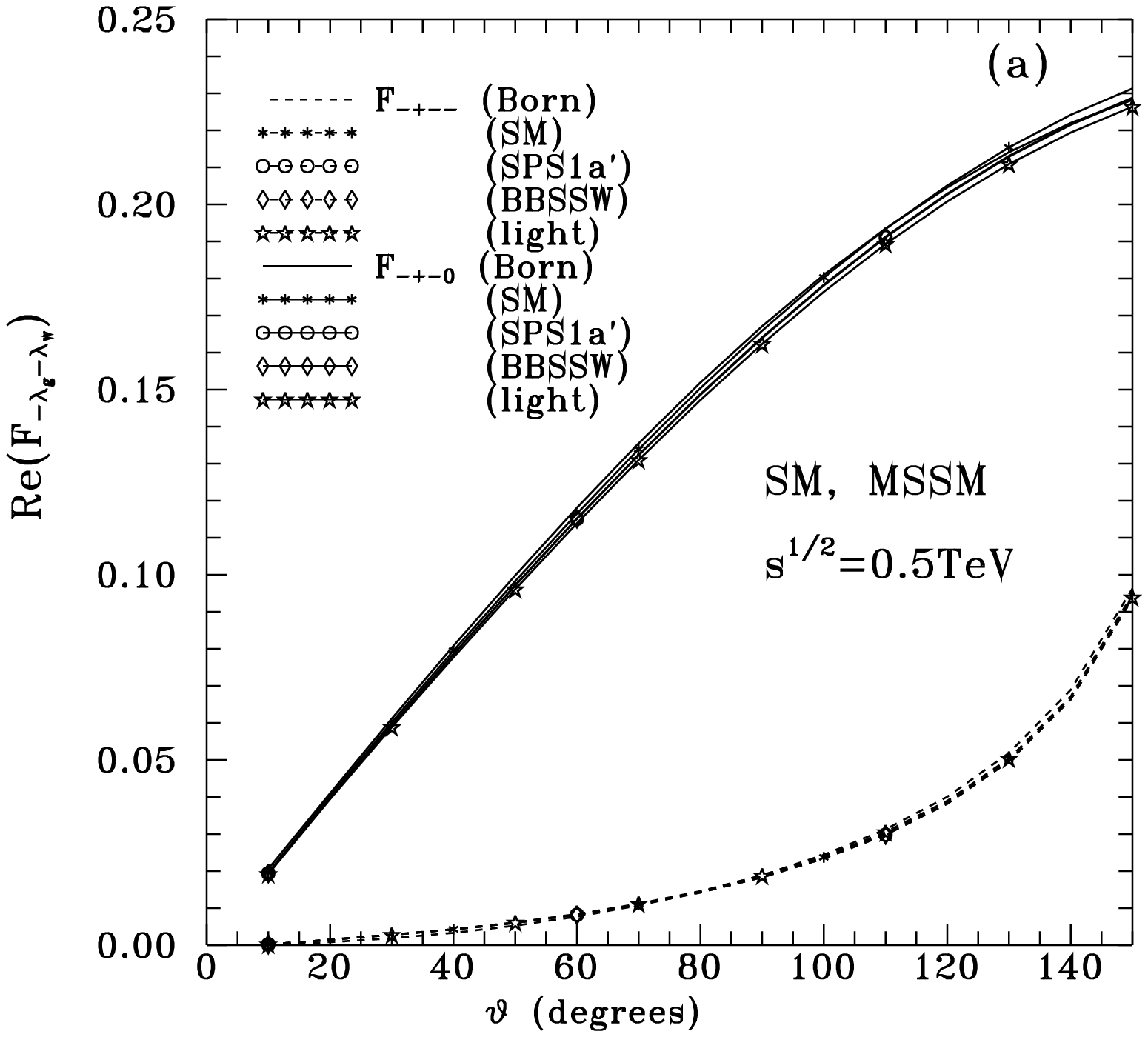,height=7.4cm}\hspace{0.5cm}
\epsfig{file=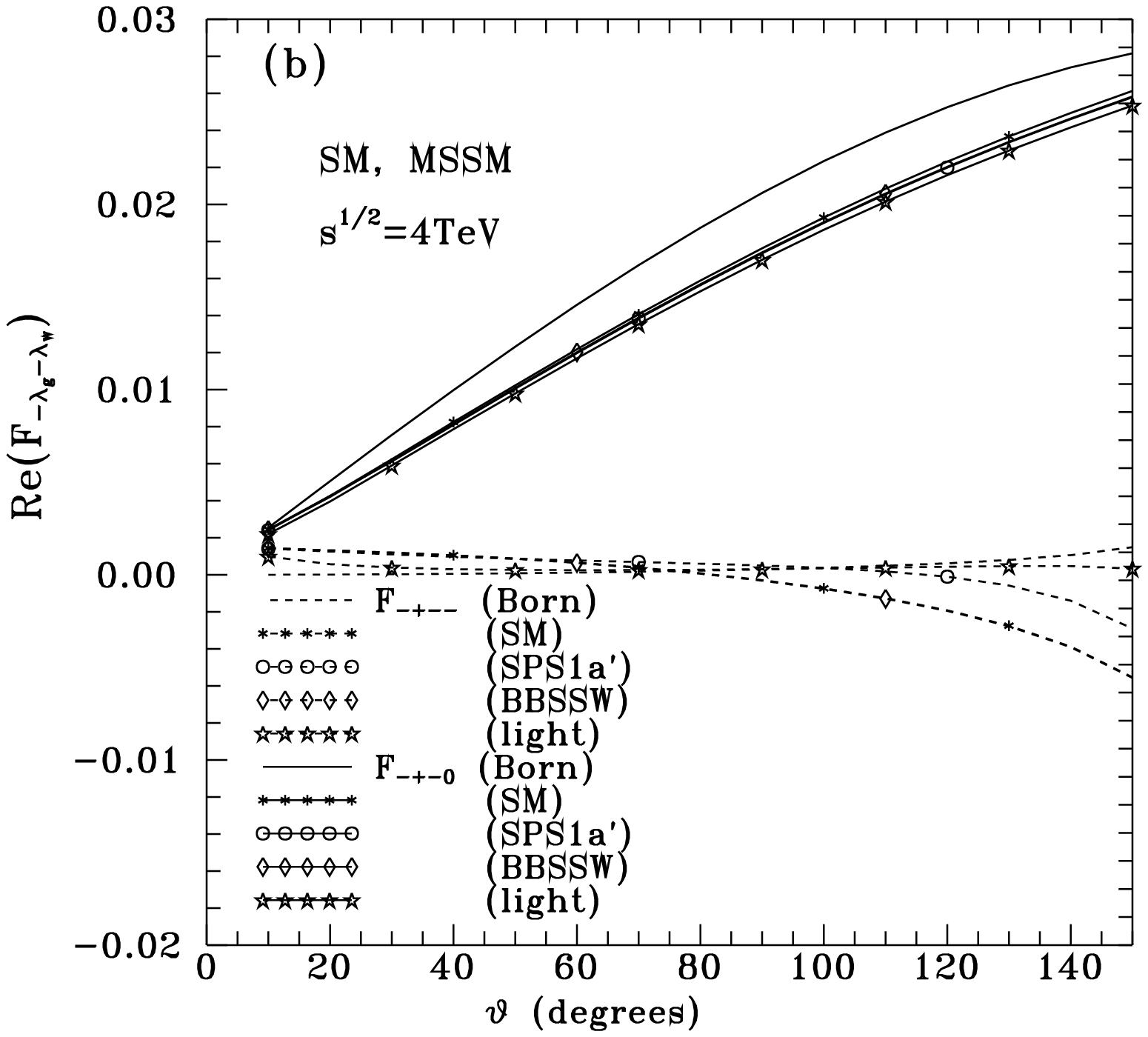,height=7.4cm}
\]
\caption[1]{Angular  dependence
of the helicity violating GBHV2 amplitudes $F_{-+--},~F_{-+-0}$
in the Born approximation, and  SM and   MSSM benchmark models, at
  c.m. energies   0.5 TeV (a), and 4 TeV (b).}
\label{HV2-angle-fig}
\end{figure}

\begin{figure}[b]
\[
\hspace{-0.5cm}
\epsfig{file=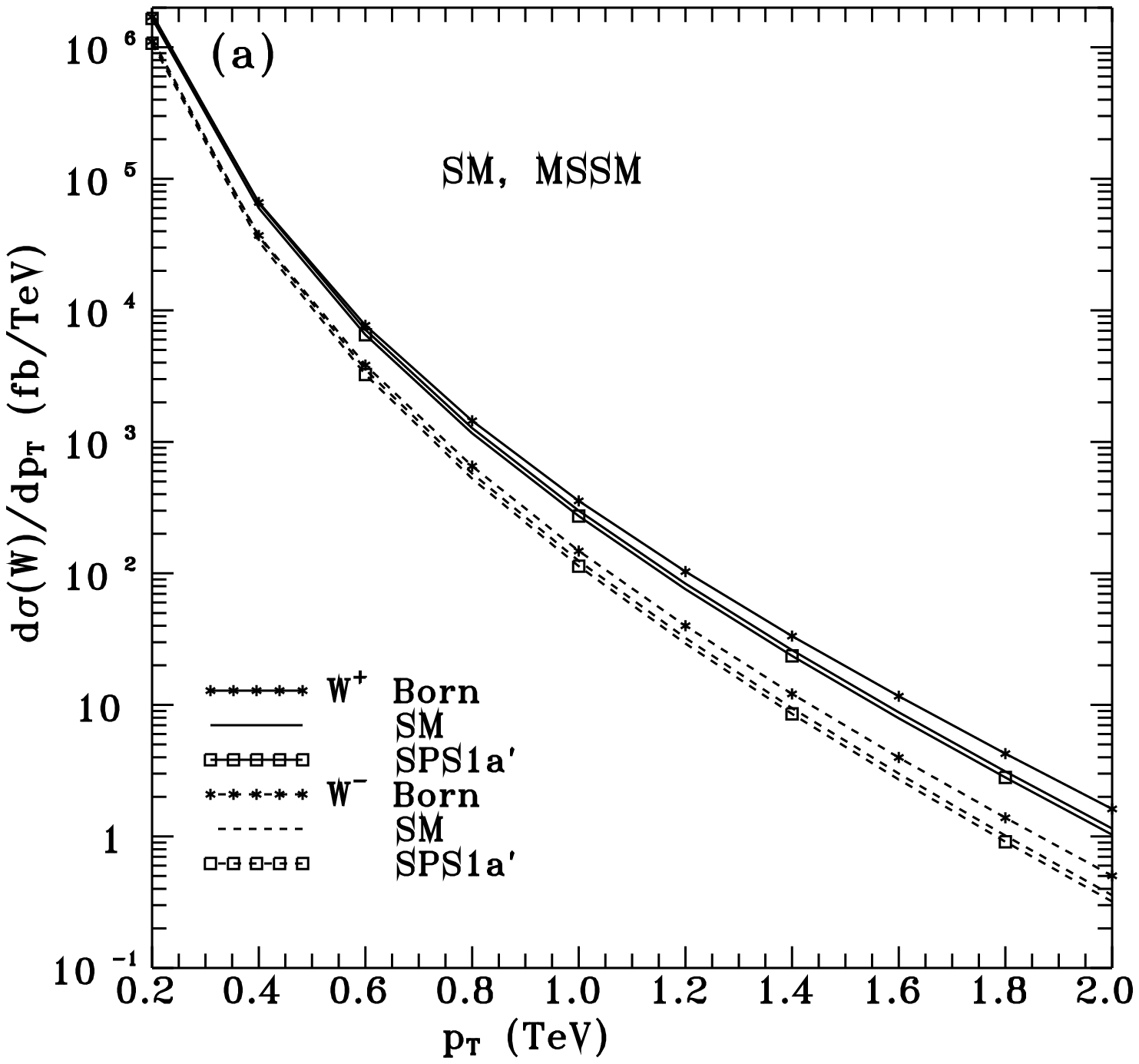,height=7.5cm}\hspace{0.5cm}
\epsfig{file=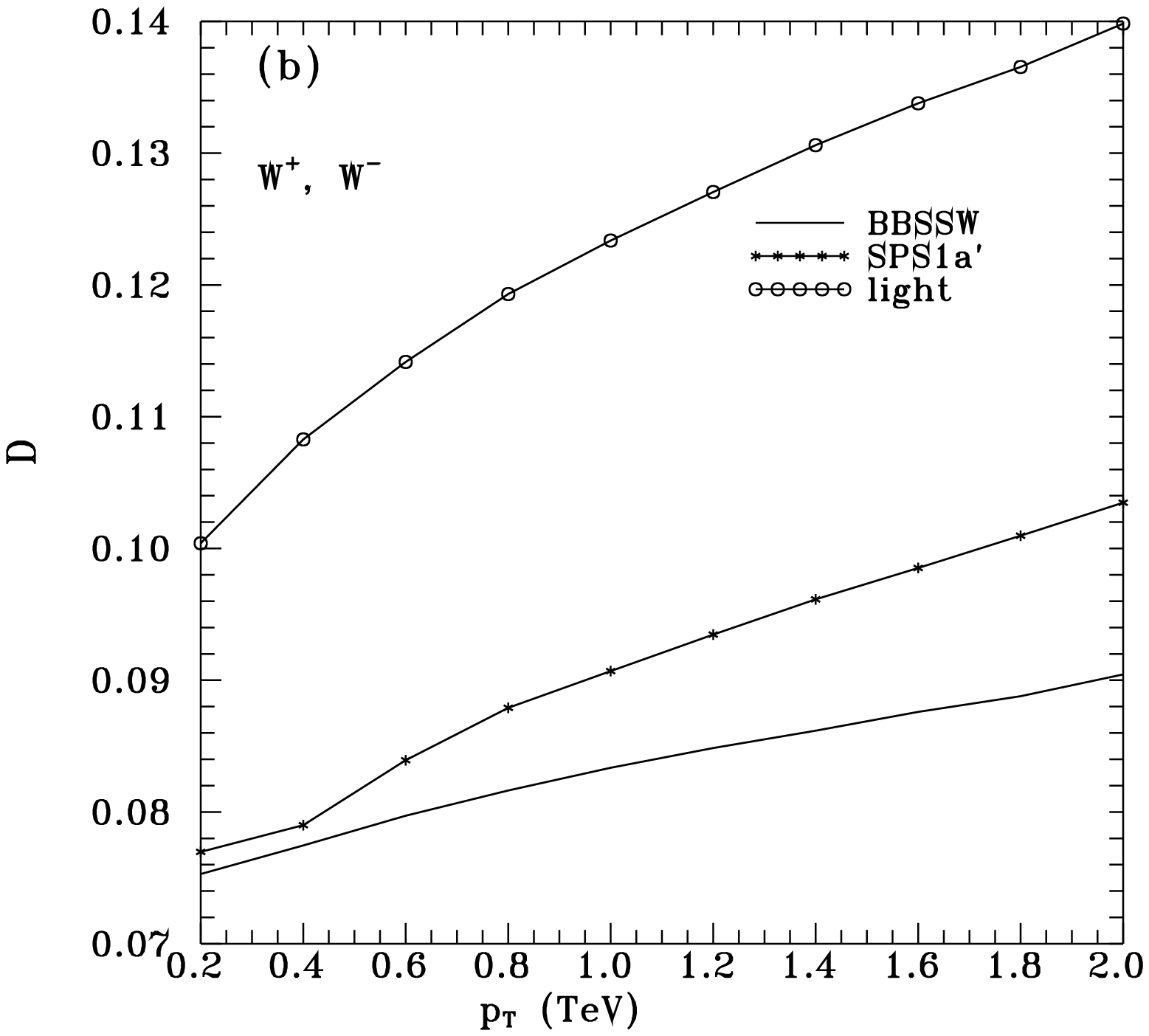,height=7.5cm}
\]
\caption[1]{(a):The $p_T$-distribution of a $W^\pm$ produced
in association with a jet  at LHC, in the Born approximation and
 the 1-loop SM and   $SPS1a'$ models.
(b) The percentage decrease for the  $W^\pm$-production
in  MSSM, as compared  to the SM predictions, using
(\ref{ratio-D}) and the MSSM models of Table 1. }
\label{W-sig-fig}
\end{figure}

\begin{figure}[p]
\vspace*{-2.cm}
\[
\hspace{-0.5cm}
\epsfig{file=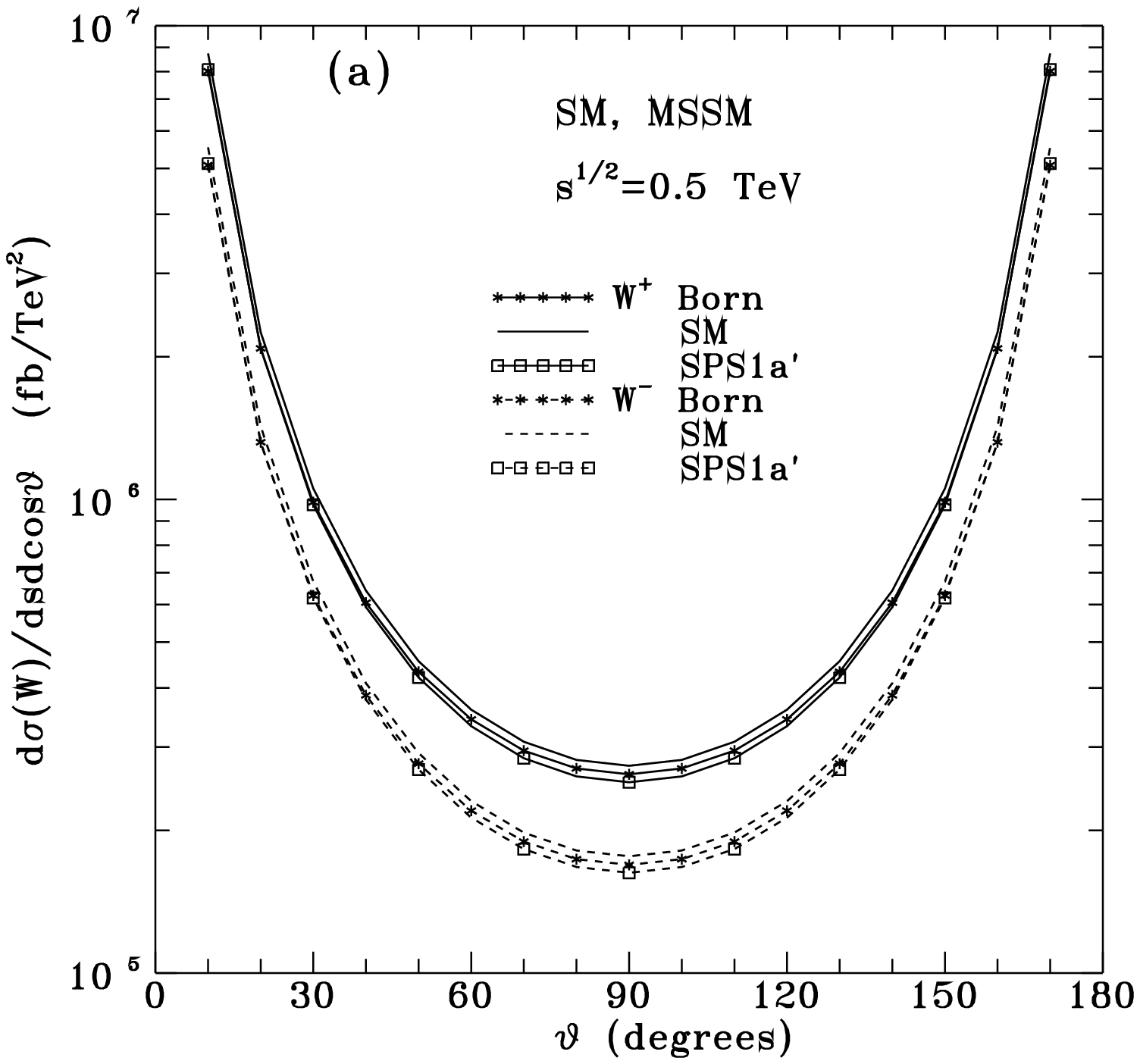,height=7.5cm}\hspace{0.5cm}
\epsfig{file=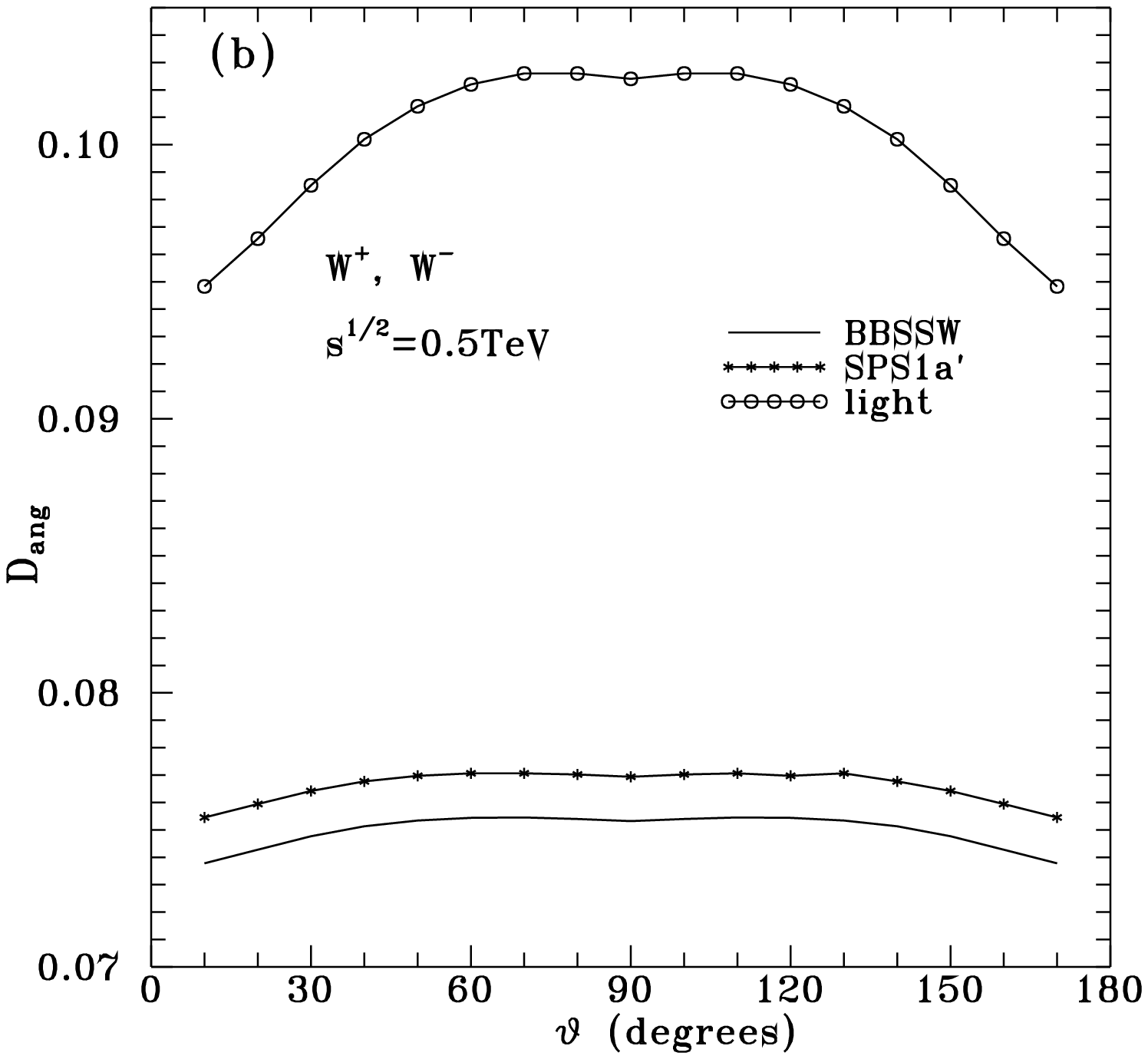,height=7.5cm}
\]
\vspace*{0.5cm}
\[
\hspace{-0.5cm}
\epsfig{file=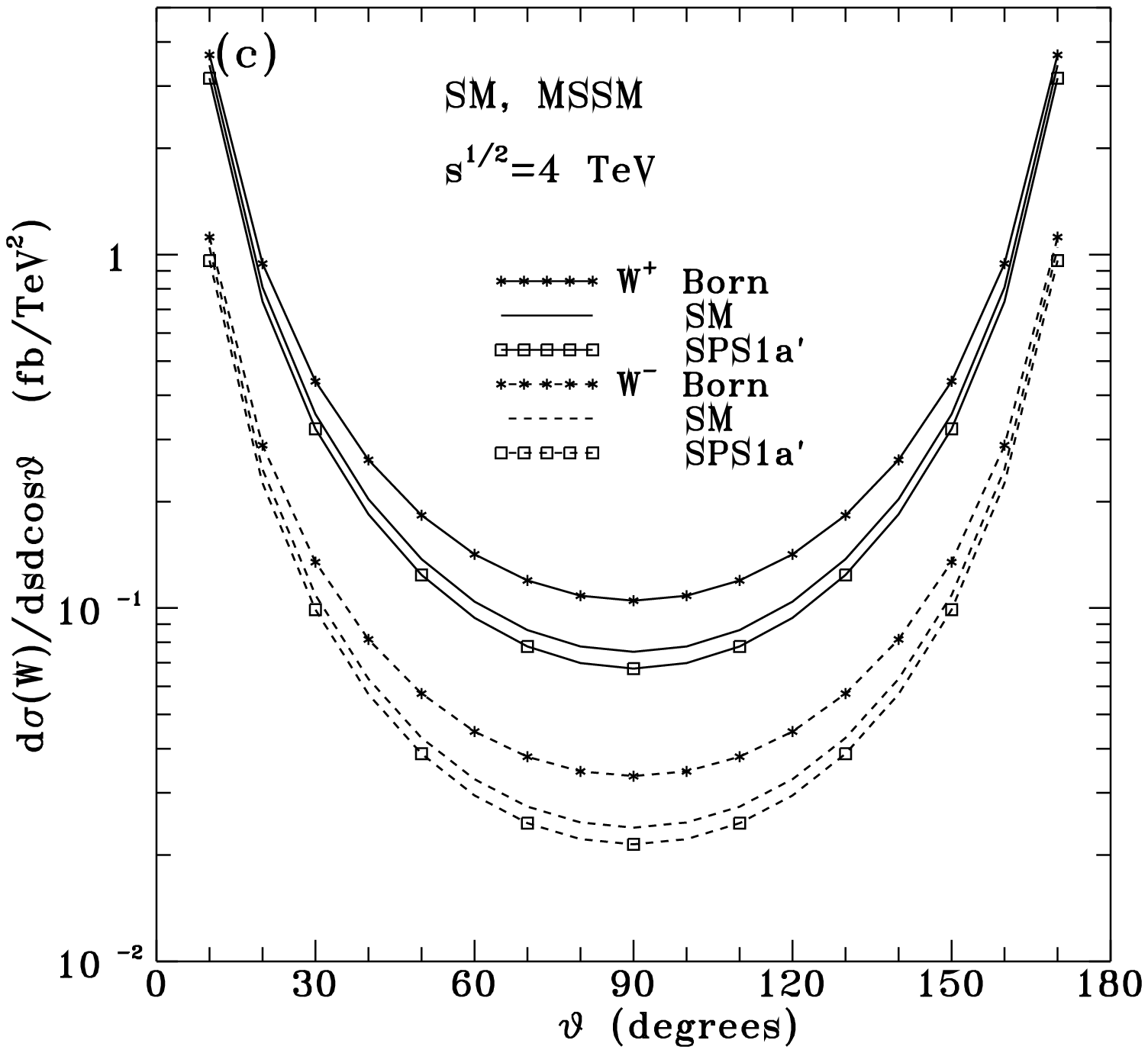,height=7.5cm}\hspace{0.5cm}
\epsfig{file=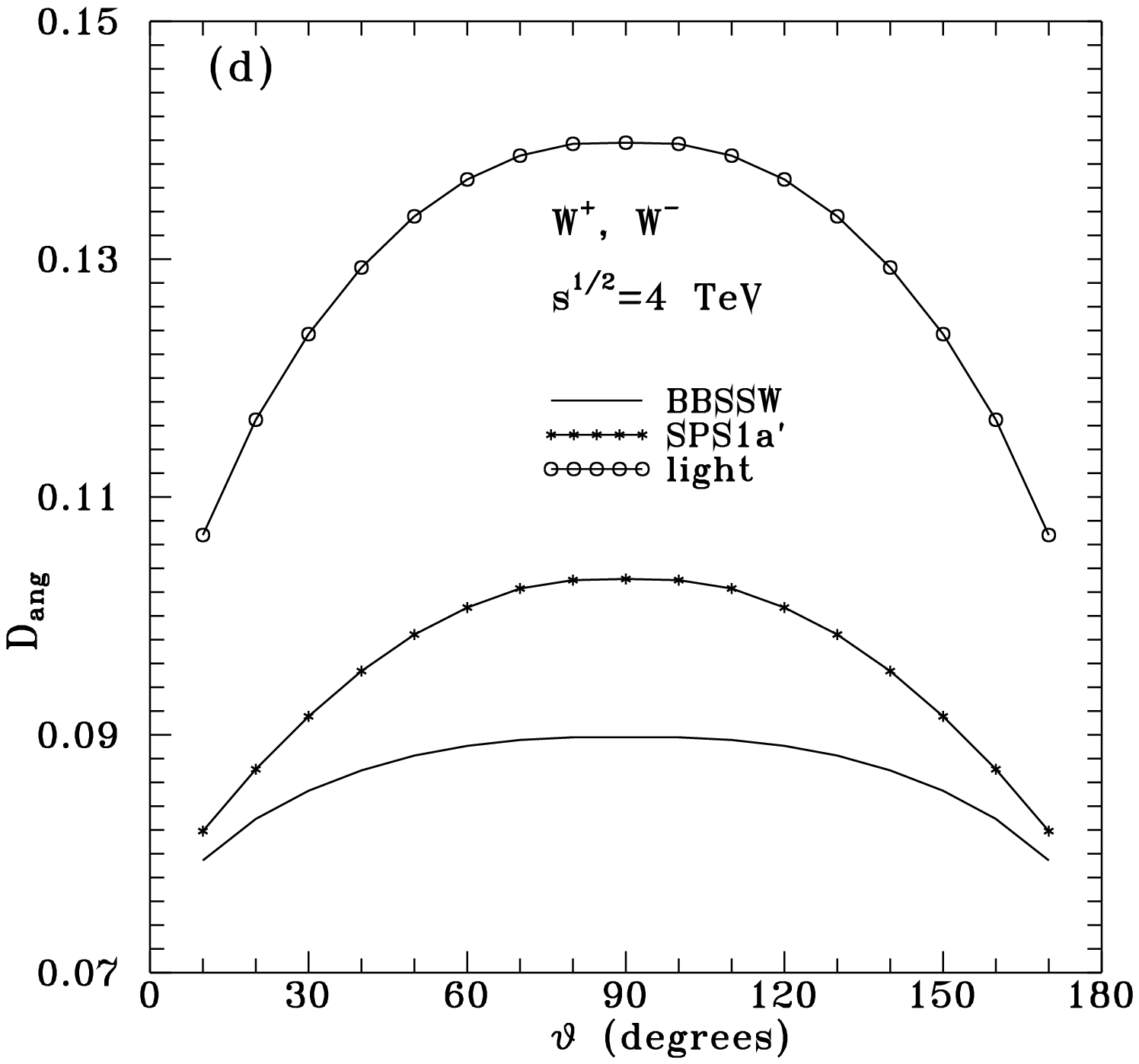,height=7.5cm}
\]
\caption[1]{The angular distribution at the subprocess c.m.,
for  subprocess energies 0.5 TeV (a), and 4 TeV (c), of a $W^\pm$ produced
in association with a jet  at LHC. The results describe the predictions in
the Born approximation and
 the 1-loop SM and   $SPS1a'$ models. In (b) and (d)
the corresponding percentage decreases are given  for the  $W^\pm$-production
in  MSSM, as compared  to the SM predictions, using
(\ref{ratio-Dang})  and the same MSSM models as in Fig.\ref{W-sig-fig}. }
\label{W-angsig-fig}
\end{figure}

In practice, the vanishing of GBHV amplitudes in MSSM,
is more or less precocious, depending on the
specific SUSY model and the value of $M_{SUSY}$.
Compare Figs.\ref{HV1-energy-fig}b,\ref{HV2-energy-fig}b.

Thus,  for  the benchmarks of  Table 1,
 the GBHC amplitudes for $ug\to dW^+$ fully dominate
the process, at energies above (2-3) TeV. Moreover, these GBHC amplitudes
can be adequately  described
by (\ref{ReFmmmm-LL},\ref{ReFmpmp-LL}) at 1-loop, in both SM and MSSM.

Particularly for MSSM, the only   model dependent parameter in these formulae
is  $M_{SUSY}$. In the  actual examples presented here, $M_{SUSY}$,
as well as the constants in Table 2,
were  estimated by comparing (\ref{ReFmmmm-LL},\ref{ReFmpmp-LL})
to the exact 1-loop result.\\

Finally, we have also presented   the global electroweak SUSY effects
arising at 1-loop at LHC, for the $W^{\pm}$ transverse
momentum distribution and the angular distribution in the c.m.
of the produced $W^\pm$+jet pair.
These effects are induced by    contributions from the subprocesses $qg\to q'W$,
$\bar q g \to\bar q' W$, $q\bar q'\to Wg$, which have been  obtained
from the basic process $ug\to dW$ through crossing.
The SUSY effect  has been
 typically found to be in the $10\%$ region, and \underline{reduces} the SM
 expectation. Such an effect  is sufficiently large to be
observable. Note that such a  negative SUSY effect actually affects
all six helicity  amplitudes.\\

In concluding this paper, we may add two  comments on the HC asymptotic rule
proved in \cite{heli}, for all two-body processes in MSSM.
Supersymmetry was crucial in establishing that all amplitudes
violating  HC,  should exactly vanish,
asymptotically.  This feature  of SUSY,
which has no direct connection  to  its ultraviolet behavior,
 seems  to  be due to the interconnections between the MSSM and
 SM spectra and couplings,  and  it certainly deserves further study.

The other interesting thing is that for
$ug\to d W^+$, the  HC theorem appears to be applicable, already
 within the LHC range.
There are several other processes,
observable at LHC, to which HC should also apply.
I would be  intriguing  to see
whether such an early HC applicability  appears for them  too.\\

\noindent{\large\bf{Acknowledgement}}\\
\noindent
G.J.G. gratefully acknowledges the support by the European Union  contracts
MRTN-CT-2004-503369 and   HEPTOOLS,  MRTN-CT-2006-035505.

\newpage

\newpage

\newpage

\newpage

\newpage

\newpage

\end{document}